\documentstyle[aps,preprint,epsf]{revtex}
\tightenlines
\begin{document}

\title{Seniority isomerism in proton-rich $N=82$ isotones
and its indication to stiffness of the $Z=64$ core}

\author{T. Matsuzawa$^1$, H. Nakada$^2$, K. Ogawa$^2$ and G. Momoki$^3$}

\address{$^1$Graduate School of Science and Technology,
Chiba University, Inage, Chiba 263-8522, Japan\\
$^2$Department of Physics, Faculty of Science, Chiba University,
Inage, Chiba 263-8522, Japan\\
$^3$College of Industrial Technology, Nihon University,
Narashino, Chiba 275-8575, Japan}

\date{\today}

\maketitle

\begin{abstract}
The $10^+$ and $27/2^-$ isomers of the $Z>64$, $N=82$ nuclei
are investigated in the shell model framework.
We derive an extended seniority reduction formula
for the relevant $E2$ transition strengths.
Based on the extended formula,
as well as on the approximate degeneracy among the $0h_{11/2}$,
$2s_{1/2}$ and $1d_{3/2}$ orbits,
we argue that the $B(E2)$ data require the $^{146}$Gd core excitation.
The energy levels of both parities and the $B(E2)$ values
are simultaneously reproduced by a multi-$j$ shell model calculation
with the MSDI, if the excitations from $(0g_{7/2}1d_{5/2})$
to $(0h_{11/2}2s_{1/2}1d_{3/2})$ are taken into account.
\end{abstract}


\section{Introduction}\label{sec:intro}

Through recent experiments on unstable nuclei,
it has been recognized that the nuclear magic numbers
are not rigorous and somewhat depend on $Z$ and $N$~\cite{ref:magic}.
The magicity observed around the $\beta$-stable line
may disappear in a region far from the stability.
For instance, the magicity of $N=8$ no longer holds
in the neutron-rich nucleus $^{11}$Be.
Although there has been no clear evidence,
it is also of interest whether new magic numbers emerge
in proton- or neutron-rich region.
So-called submagic numbers such as $Z=40$ and $Z=64$ have been known,
which have been distinguished from the magic numbers
partly because their magicity disappears as $Z$ or $N$ changes.
However, we now know that even the usual magic numbers
depend more or less on $Z$ or $N$.
A question should be recast:
what is the difference between magic numbers
and submagic numbers?
In this respect, it is worthwhile reinvestigating
the stiffness of the subshell closure.

The $^{146}$Gd nucleus shows several indications
of the $Z=64$ subshell closure ({\it e.g.} relatively high
excitation energy of $2^+_1$)~\cite{ref:146Gd}.
In the $Z>64$, $N=82$ isotones,
high-spin isomers with $J^\pi=10^+$ (for even-$Z$ nuclei)
and $27/2^-$ (for odd-$Z$ nuclei) have systematically been
observed~\cite{ref:exp1,ref:exp5,ref:exp2,ref:exp3}.
In connection to these isomers, the single-$j$ shell model
with the $\pi 0h_{11/2}$ orbit was successfully applied
to the $Z>64$, $N=82$ isotones~\cite{ref:Lawson}.
In the single-$j$ shell model,
the seniority reduction formula (SRF) is available
for the $E2$ decay strengths of the high-spin isomers.
The SRF had predicted strong hindrance
for the decay strengths of the isomers around $Z=70$,
which is in coincidence with the measured $E2$ properties
of the $10^+$ and $27/2^-$ isomers.
At a glance, this seems to indicate that the $Z=64$ subshell
is stiff enough for $^{146}$Gd to be treated as an inert core.
On the other side, the stiffness of the $Z=64$ core has been argued
so far.
For instance, by analyzing the excitation energy
of the $10^+$ state in $^{146}$Gd as well as
those in the $Z>64$ isotones,
significant pair excitation across $Z=64$
was insisted~\cite{ref:ACGP}.

In this article, we shall investigate
the $10^+$ and $27/2^-$ isomers in the $Z>64$, $N=82$ nuclei,
primarily focusing on the stiffness of the $Z=64$ core.
For the decay strengths of the isomers, we extend the SRF
so that it could apply to the multi-$j$ cases.
This formula shows that the decay strengths
reflect the stiffness of the $Z=64$ core.
If the approximate degeneracy
among the $0h_{11/2}$, $2s_{1/2}$ and $1d_{3/2}$ orbits
is taken into consideration,
the hindrance of the $E2$ strengths of the isomers
turns out to indicate the presence of the pair excitation
across $Z=64$.

\section{Single-\lowercase{$j$} shell model for $Z>64$, $N=82$ isotones}
\label{sec:single-j}

The proton-rich $N=82$ isotones
have been explored experimentally.
After the discovery of the $Z=64$ submagic nature
at $^{146}$Gd~\cite{ref:146Gd},
several low-lying levels have been established
up to $^{154}$Hf~\cite{ref:TI96}.
In this region, the excitation energies of the yrast states
are nearly constant from nucleus to nucleus,
both for even-$Z$ ($^{148}$Dy, $^{150}$Er, $^{152}$Yb and $^{154}$Hf)
and odd-$Z$ ($^{149}$Ho, $^{151}$Tm and $^{153}$Lu) isotones.
Furthermore, high-spin isomers were observed systematically;
$10^+$ isomers for the even-$Z$ isotones around $E_x\sim 3$~MeV,
and $27/2^-$ isomers for the odd-$Z$ isotones around $E_x\sim 2.5$~MeV.

Whereas state-of-the-art shell model calculations
with a realistic effective interaction have been applied
to the $Z\leq 64$, $N=82$ isotones~\cite{ref:AHS97},
there have not been many theoretical studies in the $Z>64$ region.
Lawson carried out a single-$j$ shell model calculation
with $\pi 0h_{11/2}$
on top of the $^{146}$Gd core~\cite{ref:Lawson}.
The residual interaction was empirically determined
from the experimental energy levels of $^{148}$Dy.
The levels of the $Z\geq 66$ isotones were reproduced
to a certain extent,
apart from the odd-parity levels for the even-$Z$ nuclei
and the even-parity ones for the odd-$Z$ nuclei,
which are outside the model space.
It is noted that, while the measured excitation energies
of the $2^+$, $8^+$ and $10^+$ states gradually decrease
as $Z$ increases,
this tendency is not reproduced in the single-$j$ model.

In the single-closed nuclei, it has been known that
the seniority $v$ is conserved to a good approximation.
This is true also in Lawson's results.
The $10^+$ and $27/2^-$ isomers decay via the $E2$ transition.
The $10^+$ isomers and their daughters $8^+$ have the seniority $v=2$,
which is carried by the $(0h_{11/2})^2$ configuration.
Similarly, the $27/2^-$ isomers and their daughters $23/2^-$
have $v=3$.
The $E2$ transition is usually described by a one-body operator.
The seniority reduction formula (SRF) is well-known
in the single-$j$ configuration.
By representing the $0h_{11/2}$ orbit by $\xi$,
the SRF for the seniority-conserving $E2$ transitions
gives~\cite{ref:Tal93}
\begin{equation}
\langle \xi^n\,v\,J^\pi_f||T(E2)||\xi^n\,v\,J^\pi_i\rangle=
{\Omega_\xi-n \over \Omega_\xi-v}
\langle \xi^v\,v\,J^\pi_f||T(E2)||\xi^v\,v\,J^\pi_i\rangle\,,
\label{eq:SRF}
\end{equation}
where $\Omega_\xi=j_\xi+1/2$.
In the present case $j_\xi=11/2$ and $\Omega_\xi=6$.
In Lawson's model the particle number $n$ should be $Z-64$.
Equation~(\ref{eq:SRF}) shows hindrance of the $E2$ strengths
by the factor $[(6-n)/(6-v)]^2$ when $n$ deviates from $v$.
This hindrance factor gives parabola behavior of $B(E2)$
as a function of $Z$
and leads to a remarkably long lifetime around $Z=70$,
{\it i.e.} $^{152}$Yb.
This stabilization mechanism is called {\it seniority isomerism}.

The experimental data on the $B(E2)$ values of the isomers
well fit to the parabola in the $66\leq Z\leq 70$ region.
Furthermore, both $E2$ strengths of the $10^+$ and the $27/2^-$ isomers
are described by a single effective charge ($e_{\rm eff}\sim 1.5e$).
In particular, the strong hindrance of the $E2$ transition
is actually detected for $^{152}$Yb,
with $B(E2;10^+\rightarrow 8^+)=0.9\pm 0.1~e^2{\rm fm}^4$~\cite{ref:exp3}.
In comparison with the data, the $E2$ strengths are overestimated
for $^{153}$Lu and $^{154}$Hf in the single-$j$ model
to a certain extent, as will be shown later.
This discrepancy should be attributed to
an effect of the orbits other than $0h_{11/2}$.

\section{Extension of the seniority reduction formula}
\label{sec:ExSRF}

Despite its success in predicting the seniority isomerism,
the single-$j$ model will be too simple to be realistic,
since the $\pi 0h_{11/2}$ orbit is not isolated.
In the odd-$Z$ $N=82$ isotones,
$1/2^+$ and $3/2^+$ states are present
in the vicinity of the $11/2^-$ states,
indicating the approximate degeneracy among the proton orbits
$0h_{11/2}$, $2s_{1/2}$ and $1d_{3/2}$.
A certain number of levels with opposite parities
(odd-parity levels for the even-$Z$ isotones
and even-parity ones for the odd-$Z$ isotones) are also observed
in the low energy regime,
which cannot be described in the $\pi 0h_{11/2}$ single-$j$ model.
We shall reinvestigate the $10^+$ and $27/2^-$ isomers
in the multi-$j$ shell model framework.

There is no evidence for a breakdown
of the neutron magic number $N=82$ in the low energy region,
except for a few states relevant to the octupole collectivity.
We hereafter maintain the $N=82$ inert core.
For the proton degrees-of-freedom,
the $50<Z<82$ major shell is considered at largest.

While the seniority isomerism in this region has been discussed
based on the SRF for the single-$j$ orbit $\pi 0h_{11/2}$,
in the following we show that the formula (\ref{eq:SRF})
can be extended to the multi-$j$ model space
with a simple modification.

Let us define the seniority in the multi-$j$ space
by the sum of the seniorities of each orbit, $v=\sum_j v_j$.
The seniority is expected to be a good quantum number
in single-closed nuclei, at least for their low-lying states.
In the high-spin isomers under interest,
the seniority is carried only by the $\pi 0h_{11/2}$ orbit,
to a good approximation.
In the $50<Z<82$ major shell,
$J^\pi=10^+$ with $v=2$ is uniquely formed
by the $(0h_{11/2})^2$ configuration,
and $J^\pi=27/2^-$ with $v=3$ by $(0h_{11/2})^3$.
The decays of the isomers occur via the $E2$ transition
without changing the seniority.
Within this major shell, the $8^+$ state with $v=2$,
the final state of the $10^+$ decay,
also has the $(0h_{11/2})^2$ configuration.
The $23/2^-$ state having $v=3$, the daughter of the $27/2^-$ decay,
is predominantly $(0h_{11/2})^3$.
Although this state may have an admixture
of $(0g_{7/2})^2(0h_{11/2})^1$ and
$(0g_{7/2})^1(1d_{5/2})^1(0h_{11/2})^1$,
the admixture will be small,
because these configurations need excitation across $Z=64$ by two protons.
Moreover, the remaining part consisting of $0^+$ pairs
is expected to have almost identical structure
between the isomers and their daughter states.
This is in accordance with the spherical BCS~\cite{ref:RS80} or
Talmi's generalized-seniority picture~\cite{ref:gss},
where quasiparticles are defined on top of the coherent $0^+$ pairs
distributing over the valence orbits.
Keeping this situation in mind,
we derive an extended formula in somewhat general manner.

Suppose that (a) the seniority is a good quantum number,
(b) for a seniority-conserving $E2$ transition,
the seniority is carried by a single orbit (labelled by $\xi$)
both for the initial and the final states of the transition,
and (c) the wave functions of the paired particles are identical
between the two states.
The condition (c) will be given below
in more definitive manner.
We represent all valence orbits other than $\xi$ by $r$.
In the present $N=82$ case, $\xi=\pi 0h_{11/2}$
and $r=\pi (0g_{7/2}1d_{5/2}1d_{3/2}2s_{1/2})$.
The shell model bases are decomposed into the product
of the $\xi^{n_\xi}$ and $r^{n_r}$ configurations,
where the valence particle number is given by
$n = n_\xi + n_r$.
Because of (b), the seniorities of the $\xi$ and $r$ subspaces
are $v_\xi=v$ and $v_r=0$, respectively.
The initial and final states are expanded as
\begin{equation}
  |(\xi r)^n\,v\,J_i^\pi\rangle = \sum_{n_\xi(\geq v),\alpha}
      c_{n_\xi \alpha} |\xi^{n_\xi}\,v_\xi=v\,J_i^\pi\rangle \,
      |r^{n-n_\xi}\,\alpha\,v_r=0\ 0^+\rangle\,, \label{eq:state-i}\\
\end{equation}
and
\begin{equation}
  |(\xi r)^n\,v\,J_f^\pi\rangle = \sum_{n_\xi(\geq v),\alpha}
      c_{n_\xi \alpha} |\xi^{n_\xi}\,v_\xi=v\,J_f^\pi\rangle \,
      |r^{n-n_\xi}\,\alpha\,v_r=0\ 0^+\rangle\,. \label{eq:state-f}
\end{equation}
Here $\alpha$ represents composition of the $0^+$ pairs
within the $r^{n-n_\xi}$ configuration.
For instance, $\alpha$ distinguishes
$(0g_{7/2}1d_{5/2})^{14}(1d_{3/2})^2(2s_{1/2})^2$ from
$(0g_{7/2}1d_{5/2})^{14}(1d_{3/2})^4$.
The condition (c) is defined as
the expansion coefficients ($c_{n_\xi \alpha}$) are
equal between $|J_i^\pi\rangle$ and $|J_f^\pi\rangle$.
For zero-range interactions like the SDI,
this condition results from (a) and (b), as is verified in Appendix.
The normalization yields
\begin{equation} \sum_{n_\xi,\alpha} c_{n_\xi \alpha}^2 = 1\,.
\label{eq:norm}\end{equation}
Because of the condition (c), the occupation number on the orbit $\xi$
is equal between the initial and the final states:
\begin{equation} \langle N_\xi\rangle =
\sum_{n_\xi,\alpha} c_{n_\xi \alpha}^2 n_\xi\,,
\label{eq:numb}\end{equation}
where $N_\xi$ stands for the number operator on $\xi$.

Since the $r$ subspace carries no seniority
under the condition (b),
the $E2$ transition is forbidden within this subspace.
Namely, in the seniority-conserving $E2$ transition,
the $r$ subspace behaves as a spectator.
The $E2$ matrix element is then written as
\begin{equation}
  \langle (\xi r)^n\,v\,J_f^\pi||T(E2)||(\xi r)^n\,v\,J_i^\pi\rangle
  = \sum_{n_\xi,\alpha} c_{n_\xi \alpha}^2
     \langle \xi^{n_\xi}\,v_\xi=v\,J_f^\pi||T(E2)
     ||\xi^{n_\xi}\,v_\xi=v\,J_i^\pi \rangle\,.
  \label{eq:ExSRF0}
\end{equation}
This $E2$ transition is a non-collective one,
contributed only by the $\xi$ orbit.
Substitution of the SRF for the orbit $\xi$
(see Eq.~(\ref{eq:SRF})) into the right-hand side (RHS) yields
\begin{equation}
  \langle (\xi r)^n\,v\,J_f^\pi||T(E2)||(\xi r)^n\,v\,J_i^\pi\rangle
  = \sum_{n_\xi,\alpha} c_{n_\xi \alpha}^2
     {{\Omega_\xi-n_\xi}\over{\Omega_\xi-v}}
     \langle \xi^v\,v\,J_f^\pi||T(E2)||\xi^v\,v\,J_i^\pi \rangle\,,
\label{eq:ExSRF1}
\end{equation}
with $\Omega_\xi\equiv j_\xi+1/2$.
Because of Eqs.~(\ref{eq:norm}) and (\ref{eq:numb}),
we finally obtain
\begin{equation}
  \langle (\xi r)^n\,v\,J_f^\pi||T(E2)||(\xi r)^n\,v\,J_i^\pi\rangle
  = {{\Omega_\xi-\langle N_\xi\rangle}\over{\Omega_\xi-v}}
     \langle \xi^v\,v\,J_f^\pi||T(E2)||\xi^v\,v\,J_i^\pi \rangle\,.
  \label{eq:ExSRF}
\end{equation}
Equation~(\ref{eq:ExSRF}) links the $E2$ matrix element
to $\langle N_\xi\rangle$, occupation number on the orbit $\xi$.
If the effective charge parameter in $T(E2)$ is fixed in advance,
the $E2$ matrix element is determined only from $\langle N_\xi\rangle$.
Conversely, $\langle N_\xi\rangle$ can be extracted
from the $E2$ matrix element.
What determines $\langle N_\xi\rangle$ is $c_{n_\xi \alpha}$,
which represents the configuration mixing due to the pairing correlation.
Thus the $E2$ strengths of the isomers are a pairing property,
sensitive to the mixing via the pairing interaction.

Compare the formula (\ref{eq:ExSRF}) to the SRF
for the single-$j$ orbit (\ref{eq:SRF}).
Although the multi-$j$ matrix element is under discussion,
the only difference in the RHS is that the particle number $n$
is replaced by the expectation value $\langle N_\xi\rangle$.
We shall call Eq.~(\ref{eq:ExSRF})
{\it extended seniority reduction formula} (ExSRF).
The hindrance of the transition strength occurs via the factor
$[(\Omega_\xi-\langle N_\xi\rangle)/(\Omega_\xi-v)]^2$,
in parallel to the argument in the single-$j$ case,
and extraordinarily long lifetime is expected
if $\langle N_\xi\rangle \simeq \Omega_\xi$.
The ExSRF (\ref{eq:ExSRF}) obviously contains
the single-$j$ formula (\ref{eq:SRF}) as a limiting case.
Equation~(\ref{eq:ExSRF}) reduces to Eq.~(\ref{eq:SRF})
if $c_{n_\xi \alpha}=1$ for $n_\xi=n$ and $0$ for the others.
Still the difference from the single-$j$ case should be remarked.
Even when the seniority is conserved,
there could be configuration mixing due to the pairing correlations.
While the SRF (\ref{eq:SRF}) in the single-$j$ model requires
that any mixing should be negligible,
the ExSRF (\ref{eq:ExSRF}) holds with the pairing mixing.
The present conditions to the $E2$ hindrance are thereby
much more realistic than in the single-$j$ case,
and the hindrance due to $\langle N_\xi\rangle$ may be found
in a variety of the single-closed nuclei and their neighbors.
We shall call this mechanism {\it extended seniority isomerism}.

Blomqvist suggested, without proof, that $n$ in the SRF (\ref{eq:SRF})
can be reinterpreted as the occupation number~\cite{ref:Blom84}.
Discussion based on the BCS approximation
was given in Ref.~\cite{ref:BCS}.
The BCS argument leads to the factor $(u_\xi^2 - v_\xi^2)$
in terms of the $u$- and $v$-coefficients~\cite{uv-factor},
which is proportional to $(\Omega_\xi-\langle N_\xi\rangle)$.
However, the degree of the approximation was not clear enough.
The BCS approximation presumes coherent pairing
and ignores some dependence on the seniority
({\it e.g.} the seniority-dependence
in the denominator of Eq.~(\ref{eq:ExSRF})).
On the other hand,
we have derived the ExSRF in more rigorous and general manner,
which is exact as far as the conditions (a--c) are satisfied.

We here comment on the relation of the ExSRF (\ref{eq:ExSRF})
to the multi-$j$ quasi-spin (QS) formula
for the degenerate single-particle orbits~\cite{ref:AI66}.
The multi-$j$ QS formula is available
when the pair distributes over all the valence orbits
with equal amplitudes.
We then have
\begin{equation}
  {{n_\xi-v}\over{n-v}} = {{\Omega_\xi-v}\over{\Omega-v}}\,,
\label{eq:pair}
\end{equation}
where we use the notation
\begin{equation}
  \Omega = \sum_{j\in(\xi,r)} \Omega_j = \sum_{j\in(\xi,r)} (j+1/2)\,.
\end{equation}
By employing Eqs.~(\ref{eq:norm}) and (\ref{eq:pair}),
Eq.~(\ref{eq:ExSRF1}) reduces to
\begin{equation}
  \langle (\xi r)^n\,v\,J_f^\pi||T(E2)||(\xi r)^n\,v\,J_i^\pi\rangle
  = {{\Omega-n}\over{\Omega-v}}
     \langle \xi^v\,v\,J_f^\pi||T(E2)||\xi^v\,v\,J_i^\pi \rangle\,.
  \label{eq:GSRF}
\end{equation}
Because of the condition (b),
Eq.~(\ref{eq:GSRF}) is equivalent to the multi-$j$ QS formula
\begin{equation}
  \langle (\xi r)^n\,v\,J_f^\pi||T(E2)||(\xi r)^n\,v\,J_i^\pi\rangle
  = {{\Omega-n}\over{\Omega-v}}
     \langle (\xi r)^v\,v\,J_f^\pi||T(E2)||(\xi r)^v\,v\,J_i^\pi \rangle\,.
  \label{eq:GSRF0}
\end{equation}

We now return to the case of the $N=82$ isotones.
Since $\xi=0h_{11/2}$ (thereby $\Omega_\xi=6$),
the $10^+$ or $27/2^-$ isomer has remarkably long life
for a nucleus satisfying $\langle N_{0h_{11/2}}\rangle \simeq 6$.
Namely, the observed long lifetime of the $10^+$ isomer in $^{152}$Yb
implies $\langle N_{0h_{11/2}}\rangle \simeq 6$.
As far as $2s_{1/2}$ and $1d_{3/2}$ lie closely to $0h_{11/2}$,
there should be mixing among these orbits due to the pairing interaction,
causing decrease of $\langle N_{0h_{11/2}}\rangle$.
However, it can be compensated by the excitation
from $0g_{7/2}$ or $1d_{5/2}$ to $0h_{11/2}$,
which increases $\langle N_{0h_{11/2}}\rangle$.
As we shall discuss in the following sections,
this should be what happens in the isomers in the $Z>64$, $N=82$ isotones.

\section{Multi-\lowercase{$j$} shell model
for $Z\protect\agt 64$, $N=82$ isotones}
\label{sec:multi-j}

\subsection{Model space}
\label{subsec:space}

In this section, we present how the properties of the high-spin isomers
are described,
by a calculation in the multi-$j$ shell model framework.
As discussed in the preceding section,
the model space should include all the five orbits
in the $50<Z<82$ major shell.
Large-scale shell model calculations were carried out
for the $62\leq Z\leq 65$, $N=82$ isotones
with moderate truncation~\cite{ref:EJV99},
as well as for the $Z\leq 64$, $N=82$ isotones
in the full major shell~\cite{ref:AHS97}.
On the other hand,
our main purpose is to illustrate the extended seniority isomerism
in the $Z>64$, $N=82$ isotones.
In order to avoid time-consuming computations,
we adopt relatively small space by truncation.

The space for diagonalization is truncated as follows.
Partially maintaining the $Z=64$ subshell structure,
we restrict the excitation out of the $0g_{7/2}$ and $1d_{5/2}$ orbits
to four particles.
Furthermore, the total seniority is limited to $v\leq 3$ ($v\leq2$)
for the odd-$Z$ (even-$Z$) nuclei.
The seniorities of the $10^+$ and the $27/2^-$ states
are pure in this space as well as those of their decay daughters;
the condition (a) in Section~\ref{sec:ExSRF} is satisfied.
The condition (b) is exact for the $10^+$ decay,
while the final state $23/2^-$ of the $27/2^-$ decay
has a small admixture of the $(0g_{7/2})^2(0h_{11/2})^1$ and
$(0g_{7/2})^1(1d_{5/2})^1(0h_{11/2})^1$ configurations,
as stated in Section~\ref{sec:ExSRF}.

\subsection{Energy levels}
\label{subsec:energy}

The shell model Hamiltonian is written as
\begin{equation}
H = E_0 + \sum_j \epsilon_j N_j + V .
\label{eq:sh-H}
\end{equation}
Here $E_0$ is a constant shifting the origin of the energy,
$\epsilon_j$ represents the single-particle energy of the orbit $j$,
and $N_j$ the number operator on $j$.
The residual two-body interaction is denoted by $V$,
for which we adopt the modified surface-delta interaction (MSDI),
\begin{eqnarray}
V = -4\pi A_{T=1} \sum_\lambda Y^{(\lambda)}(\hat{\mathbf r}_1)
\cdot Y^{(\lambda)}(\hat{\mathbf r}_2) + B.
\label{eq:MSDI}
\end{eqnarray}
There are 8 parameters in the Hamiltonian,
$E_0$, $\epsilon_j$ for the five orbits, $A_{T=1}$ and $B$.
They can be classified into two groups.
One is comprised of the differences of $\epsilon_j$'s (4 parameters)
and $A_{T=1}$.
These five parameters are relevant to the excitation spectra
for an individual nucleus.
The other consists of $E_0$, $B$ and overall shift of $\epsilon_j$'s.
They do not change excitation spectra, but affect the gross behavior
of the binding energies.
It is noticed that effects of the Coulomb repulsion
between protons are principally contained in $B$.
As is proven in Appendix,
the ExSRF (\ref{eq:ExSRF}) becomes exact for the $10^+$ decay
with the present seniority-truncated model space and the interaction.

\begin{figure}[h]
\epsfysize=8cm
\centerline{\epsffile{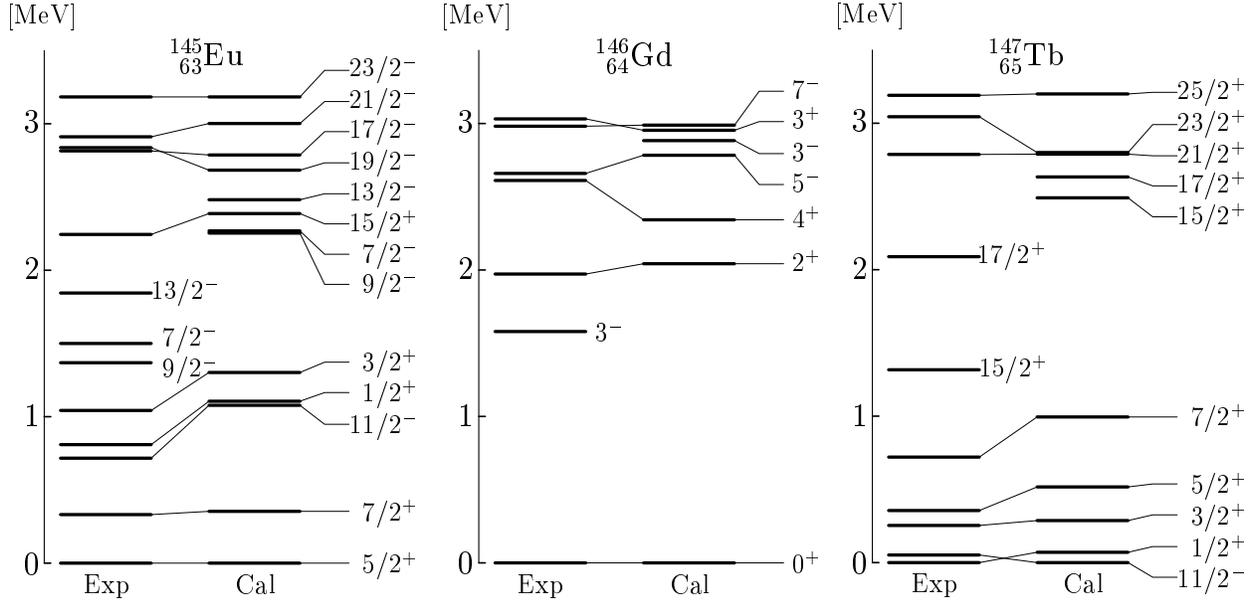}}
  \vspace{1cm}
  \caption{Comparison of the observed and calculated energy levels
  for $^{145}$Eu, $^{146}$Gd and $^{147}$Tb.
  The levels of $^{146}$Gd and the lowest 5 levels of
  $^{145}$Eu and $^{147}$Tb are employed to fit the parameters
  in the shell model Hamiltonian.
  The experimental data are taken from Ref.~\protect\cite{ref:TI96}.}
  \label{fig:Efit}
\end{figure}

In describing the extended seniority isomerism,
it is important to reproduce the degree of the pair excitation
out of the $Z=64$ core.
In $^{147}$Tb and $^{149}$Ho, a $5/2^+$ and a $7/2^+$ levels
have been observed at very low energies ($E_x\alt 1$~MeV)~\cite{ref:TI96}.
These levels could be another manifestation of the core excitation.
It is hard to reproduce these levels
without including the $0g_{7/2}$ and $1d_{5/2}$ orbits.
Analogously, $^{145}$Eu has low-lying ($E_x\alt 1$~MeV) states
with $11/2^-$, $1/2^+$ and $3/2^+$.
The coupling constant $A_{T=1}$ and the $\epsilon_j$ differences
are determined so as to reproduce the lowest levels
of $E_x\alt 1$~MeV in $^{145}$Eu and $^{147}$Tb,
as well as the $E_x\alt 3$~MeV low-lying levels of $^{146}$Gd.
The adopted value of $A_{T=1}$ is 0.210~MeV.
The results of the fitting are depicted in Fig.~\ref{fig:Efit},
together with several higher-lying levels,
in comparison with the experimental data.

There are a few levels which are not described by the calculation.
The $3^-$ state of $^{146}$Gd has been interpreted
as an octupole collective mode
including the neutron excitations~\cite{ref:exp1}.
Therefore this state has been excluded from the fitting.
The $9/2^-$, $7/2^-$ and $13/2^-$ states of $^{145}$Eu
are considered to be $\pi d_{5/2}^{-1}\otimes 3^-$
or $\pi g_{7/2}^{-1}\otimes 3^-$~\cite{ref:AHS97}.
Since they involve the octupole collective excitation,
these states are beyond the model space in the present calculation
as well.
The $15/2^+$ and $17/2^+$ states of $^{147}$Tb
are also regarded as $\pi h_{11/2}\otimes 3^{-}$.

\begin{figure}
\epsfysize=8cm
\centerline{\epsffile{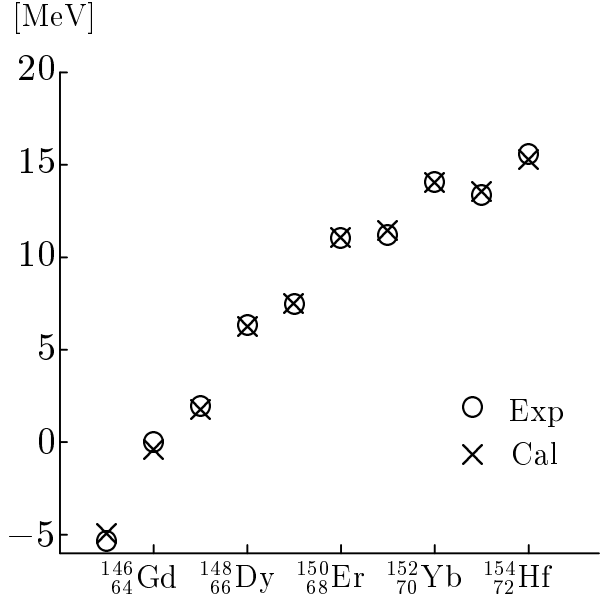}}
  \vspace{1cm}
  \caption{Binding energies for $63\leq Z\leq 74$,  $N=82$ isotones.
  All values are plotted relative to the experimental binding energy
  at $^{146}$Gd.
  The data are taken from Ref.~\protect\cite{ref:AW95}.}
  \label{fig:BE}
\end{figure}

The remaining parameters, $E_{0}$, $B$ and the constant shift of
the single-particle energies, are fixed
from the binding energies of the $63\leq Z\leq 74$,
$N=82$ isotones~\cite{ref:AW95}.
We obtain $E_0=90.10$~MeV and $B=0.409$~MeV,
representing the energies by relative values to
the experimental ground state energy of $^{146}$Gd.
The resultant single-particle energies are listed
in Table~\ref{tab:spe}.
The calculated binding energies are compared with the data
in Fig.~\ref{fig:BE}.
We have sufficiently good agreement, with the largest discrepancy
of 0.45~MeV for $^{145}$Eu.

The ground-state wave function of $^{146}$Gd holds the $Z=64$ closure
only by 11\% in this calculation.
The core is broken due to the pairing correlation,
with keeping the seniority a good quantum number.
Having 53\% excitation of a single pair
and 36\% of two pairs, average number of protons
excited out of the $Z=64$ core amounts to 2.5.
This result is barely influenced
even if we relax the seniority truncation to $v\leq 4$.
As was pointed out in Ref.~\cite{ref:ACGP},
the $Z=64$ core is broken to a sizable extent by the pair excitation.

\begin{figure}[h]
\epsfysize=8cm
\centerline{\epsffile{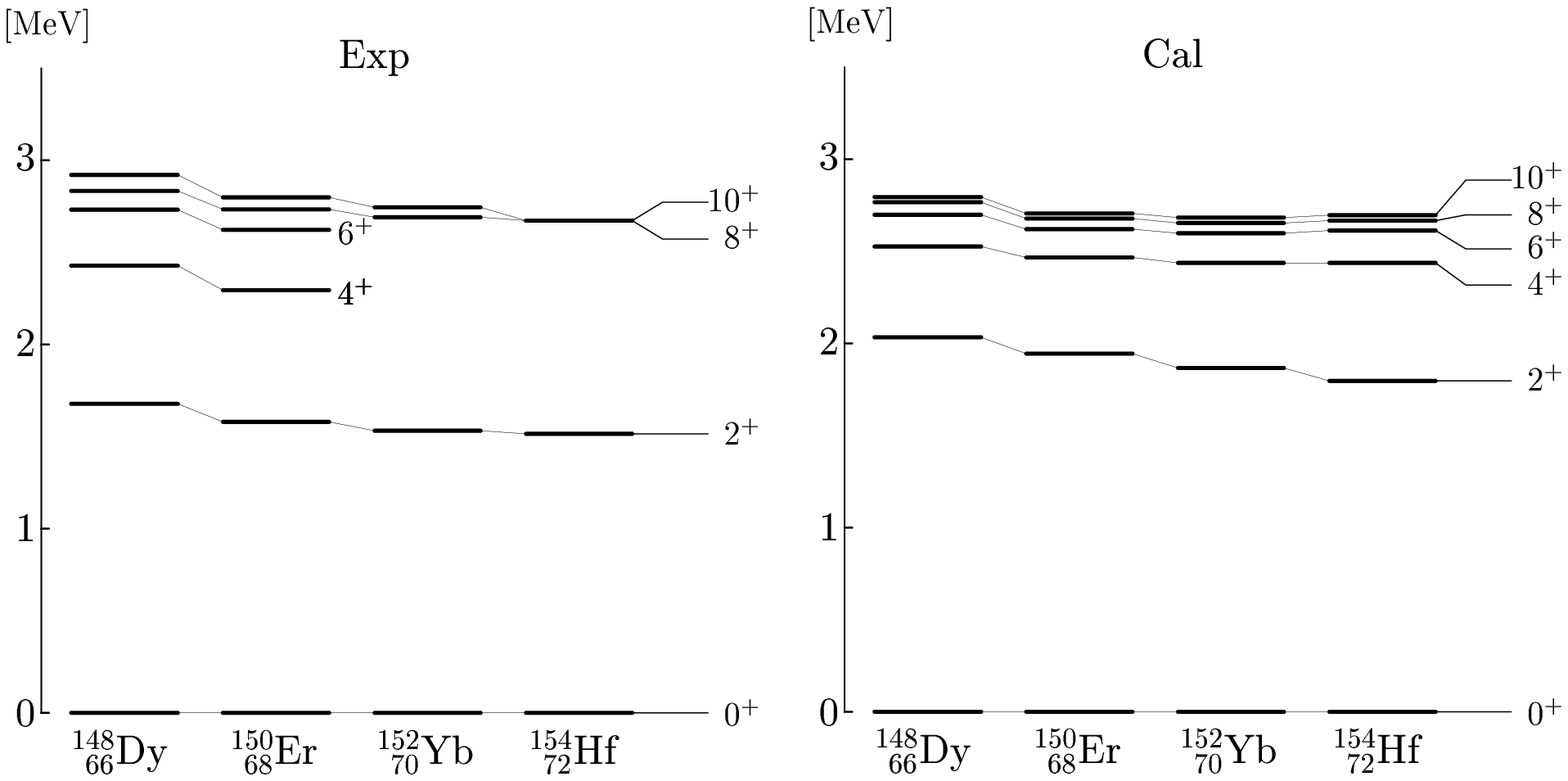}}
  \vspace{1cm}
  \caption{Comparison between the experimental and calculated
  even-parity energy levels for even-$Z$, $N=82$ nuclei.
  The experimental data are taken from Ref.~\protect\cite{ref:TI96}.}
  \label{fig:Eng_e+}
\end{figure}

\begin{figure}[h]
\epsfysize=8cm
\centerline{\epsffile{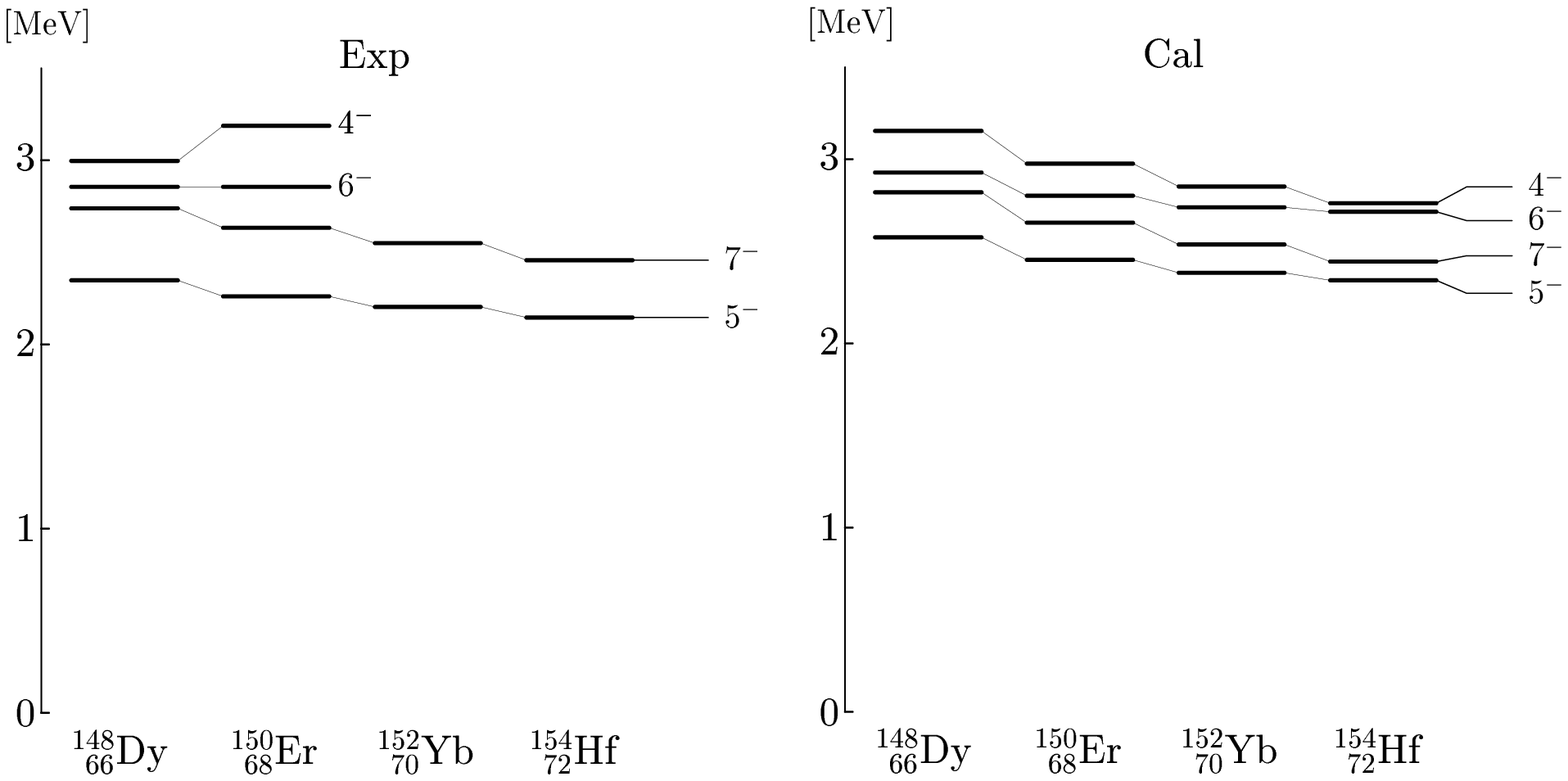}}
  \vspace{1cm}
  \caption{Comparison between the experimental and calculated
  odd-parity energy levels for even-$Z$, $N=82$ nuclei.
  The experimental data are taken from Ref.~\protect\cite{ref:TI96}.}
  \label{fig:Eng_e-}
\end{figure}

We carry out a shell model calculation with the above Hamiltonian
for the $66\leq Z\leq 72$, $N=82$ nuclei.
The calculated energy levels for the even-$Z$ nuclei
are compared with the observed ones~\cite{ref:TI96}
in Figs.~\ref{fig:Eng_e+} (for even-parity levels)
and \ref{fig:Eng_e-} (for odd-parity levels),
up to $E_x\simeq 3$~MeV.
Almost all levels in this energy range
are in reasonably good agreement.
Among them, the $E_x(2^+)$ values are somewhat higher than the data.
This discrepancy seems mainly concerned with the quadrupole collectivity,
and could be ascribed to the truncated model space
or to the interaction which might be too simple.
On the other hand,
the $E2$ decay of the isomers has non-collective character,
occurring via the transition within $0h_{11/2}$.
Therefore it is not quite relevant to the quadrupole collectivity.
As presented in Fig.~\ref{fig:Eng_e-},
the odd-parity levels are also reproduced,
except for the octupole collective state $3^-$,
which is not shown in the figure.
This is an obvious advantage
over against the previous single-$j$ calculation,
since the odd-parity levels are out of the model space
in the single-$j$ calculation.
As mentioned in Section~\ref{sec:single-j},
the excitation energies of $2^+$, $8^+$ and $10^+$ states
slightly decrease as $Z$ goes up.
This behavior is well reproduced by the present calculation,
while the energies slightly increase in the single-$j$ model.

\begin{figure}[h]
\epsfysize=8cm
\centerline{\epsffile{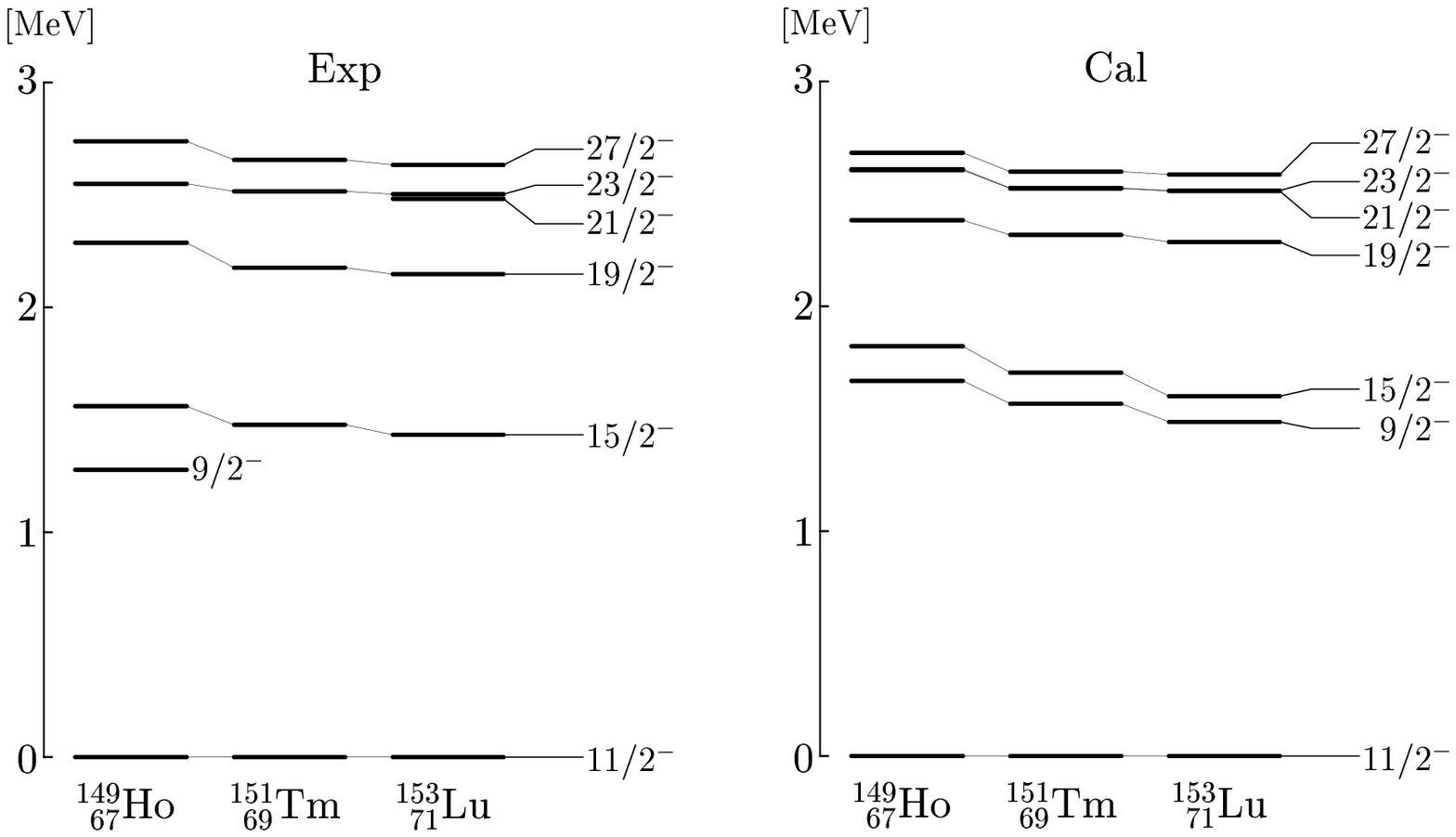}}
  \vspace{1cm}
  \caption{Comparison between the experimental and calculated
  odd-parity energy levels relative to $11/2^-$,
  for odd-$Z$, $N=82$ nuclei.
  The experimental data are taken from Ref.~\protect\cite{ref:TI96}.}
  \label{fig:Eng_o-}
\end{figure}

\begin{figure}[h]
\epsfysize=8cm
\centerline{\epsffile{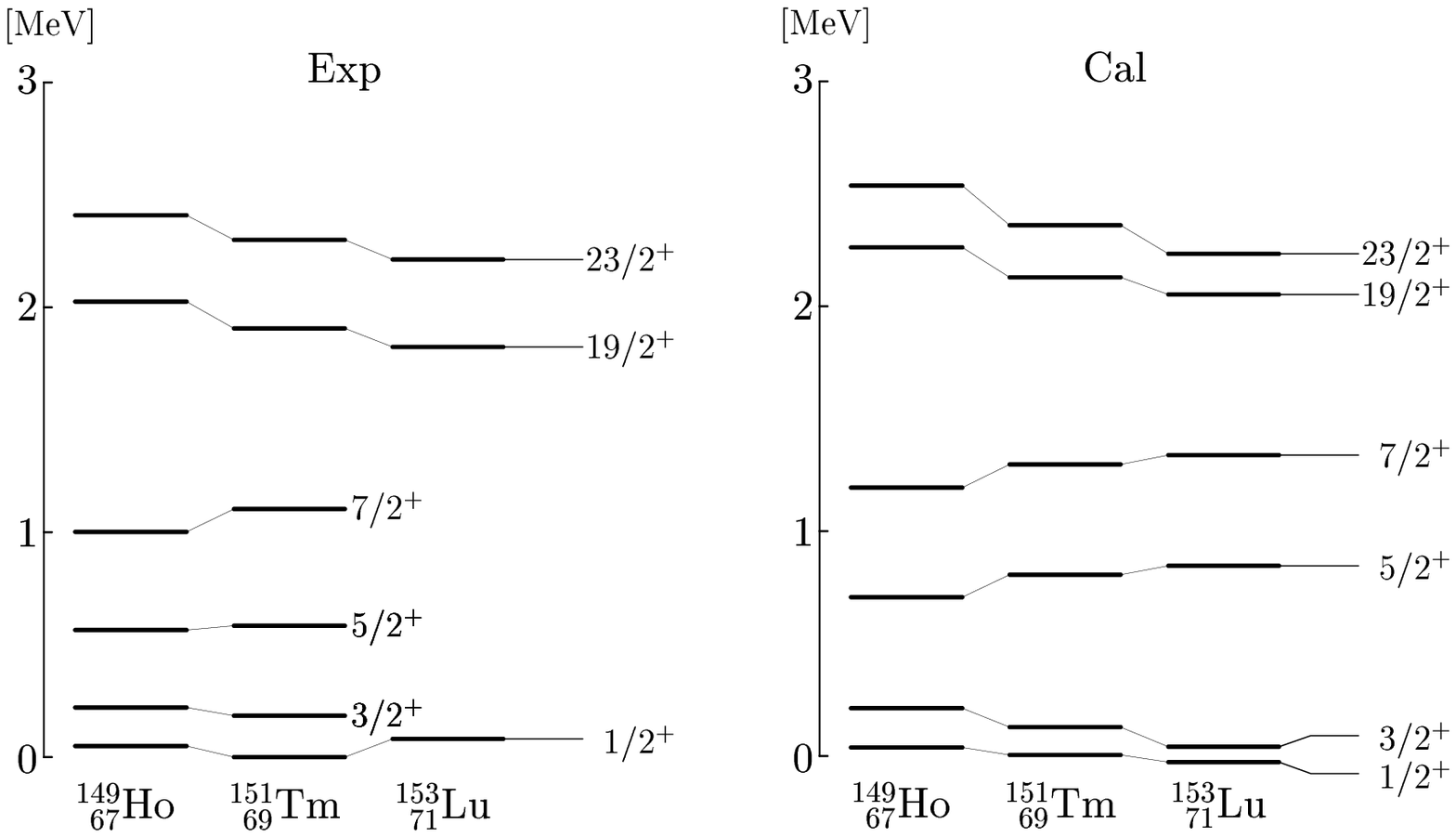}}
  \vspace{1cm}
  \caption{Comparison between the experimental and calculated
  even-parity energy levels relative to $11/2^-$,
  for odd-$Z$, $N=82$ nuclei.
  The experimental data are taken from Ref.~\protect\cite{ref:TI96}.}
  \label{fig:Eng_o+}
\end{figure}

In Figs.~\ref{fig:Eng_o-} and \ref{fig:Eng_o+},
the calculated yrast levels are compared with
the experimental data for the odd-$Z$ nuclei,
up to $E_x\simeq$3~MeV.
The energies relative to the $11/2^-$ state are presented
both for the data and the calculated results.
The agreement is sufficiently good,
as in the even-$Z$ nuclei.
The even-parity states, which are beyond the space
in the single-$j$ model with $0h_{11/2}$,
are also reproduced (Fig.~\ref{fig:Eng_o+}).
In all of the calculated levels presented in the figures,
the seniority is conserved to an excellent extent.
The $11/2^-$, $1/2^+$, $3/2^+$, $5/2^+$ and $7/2^+$ states
lying in $E_x\alt 1$~MeV have $v=1$,
while the others have $v=3$.
It should be remarked that the $5/2^+$ and $7/2^+$ levels,
the $1d_{5/2}$ and $0g_{7/2}$ states with the pair excitation,
are also reproduced well, in $^{149}$Ho and $^{151}$Tm.
The intruder level $15/2^+$ is not shown in the figure,
which should be an octupole collective state with
$\pi 0h_{11/2}\otimes 3^{-}$.

\subsection{$E2$ strengths of the high-spin isomers}
\label{subsec:E2}

Let us turn to the $E2$ transition strengths
of the high-spin isomers.
The $E2$ operator is given by
\begin{equation}
T(E2) = e_{\rm eff} \sum_{j,j'} {1\over \sqrt{5}}
  \langle j'||r^2 Y^{(2)}(\hat{\mathbf r}) ||j\rangle\,
  [a_{j'}^\dagger \tilde a_j]^{(2)}\,,
\label{eq:E2}
\end{equation}
where $\tilde a_{jm} = (-)^{j+m}a_{j-m}$.
The single-particle matrix element
$\langle j'||r^2 Y^{(2)}(\hat{\mathbf r}) ||j\rangle$
is evaluated by using the harmonic oscillator single-particle
wave functions with the oscillator parameter
$\nu (=1/b^2) = M\omega/\hbar = 0.98A^{-1/3}~{\rm fm}^{-2}$.

It should be noticed that, in the $E2$ calculation,
there remains only a single adjustable parameter $e_{\rm eff}$,
the effective charge.
It is found that $e_{\rm eff}=2.3e$
fits well to all of the $10^+$ and $27/2^-$ decays.
This value is significantly larger than the effective charge of $1.5e$
which was adopted in the single-$j$ calculation~\cite{ref:Lawson}.
This is in contrast to the collective transitions,
where $e_{\rm eff}$ should be smaller as the model space is extended,
since the matrix elements of $T(E2)/e_{\rm eff}$ tend to increase.
For the $E2$ transitions of the isomers,
which do not have collective character,
the matrix elements of $T(E2)/e_{\rm eff}$ are smaller
in the multi-$j$ case than in the single-$j$ case
at $^{148}$Dy and $^{149}$Ho,
as is recognized from Eq.~(\ref{eq:ExSRF}).

While the effective charge of $1.4e$ was recommended
in realistic calculations in the $Z<64$ region~\cite{ref:AHS97},
several calculations in the $Z\geq 64$ region
assumed $e_{\rm eff}= (2.0\sim 2.25)e$~\cite{ref:EJV99,ref:DL92}.
The origin of the difference in the effective charge
between $Z<64$ and $Z\geq 64$ is not clear,
because in either case the model space consists of
all the five orbits in the $50<Z<82$ shell,
and the orbital-dependence of the effective charge
is normally weak~\cite{ref:SG84}.
We just point out that our value seems consistent
with those of the previous studies in the $Z\geq 64$ region.

\begin{figure}[h]
\epsfysize=8cm
\centerline{\epsffile{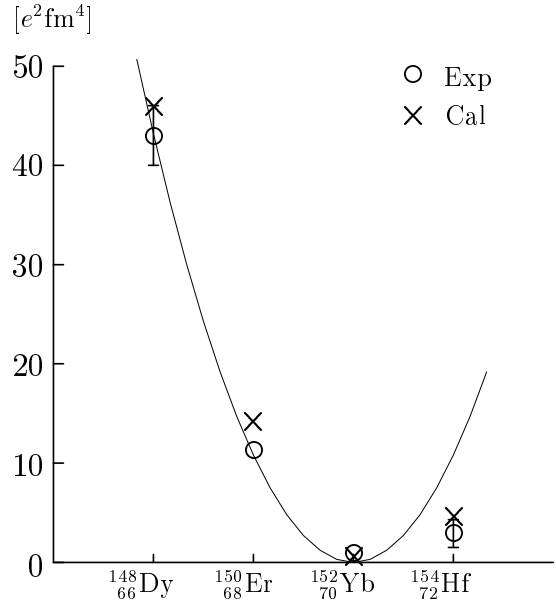}}
  \vspace{1cm}
  \caption{$B(E2;10^+\rightarrow 8^+)$ for even-$Z$, $N=82$ nuclei.
  The crosses show the results of the present calculation,
  while the thin solid line those of the $\pi 0h_{11/2}$ single-$j$
  calculation of Ref.~\protect\cite{ref:Lawson}.
  The circles stand for the experimental data
  taken from Ref.~\protect\cite{ref:exp1} for $^{148}$Dy,
  Ref.~\protect\cite{ref:exp2} for $^{150}$Er
  and Ref.~\protect\cite{ref:exp3} for $^{152}$Yb and $^{154}$Hf.
  }\label{fig:E2_e}
\end{figure}

In Fig.~\ref{fig:E2_e}, we show
the $B(E2;10^+\rightarrow 8^+)$ values
for the $66\leq Z\leq72, N=82$ isotones.
The calculated values are compared with the measured ones,
as well as with those obtained in the single-$j$ calculation
by Lawson~\cite{ref:Lawson}.
The $E2$ hindrance at $Z=70$ ({\it i.e.} $^{152}$Yb)
occurs also in the present multi-$j$ calculation.
Our calculation gives $B(E2)=0.6~e^2{\rm fm}^4$,
in good agreement with the data
$0.9\pm0.1~e^2{\rm fm}^4$~\cite{ref:exp3}.

As has been shown by the ExSRF (\ref{eq:ExSRF}),
the $E2$ strengths of the $10^+$ states are
essentially determined from the occupation number
$\langle N_{0h_{11/2}}\rangle$.
In the present calculation, the wave functions of the $10^+$ state
and of $8^+$ yield $\langle N_{0h_{11/2}}\rangle = 5.7$ in $^{152}$Yb.
This occupation number close to $\Omega_{0h_{11/2}}=6$
gives rise to the strong $E2$ hindrance.
We view this hindrance from another standpoint.
See Eq.~(\ref{eq:ExSRF0}), recalling $\xi=0h_{11/2}$.
By decomposing the wave functions as in Eqs.~(\ref{eq:state-i})
and (\ref{eq:state-f}),
we look into the contribution of each $n_\xi$ component.
Table~\ref{tab:wf_Yb} illustrates $\sum_\alpha c_{n_\xi \alpha}^2$,
$\langle \xi^{n_\xi}\,v_\xi=v\,J_f^\pi||T(E2)
||\xi^{n_\xi}\,v_\xi=v\,J_i^\pi\rangle$
and their product, for each $n_\xi(=2,4,6,8,10)$.
As the SRF tells us, the matrix element
$\langle \xi^{n_\xi}\,v_\xi=v\,J_f^\pi||T(E2)
||\xi^{n_\xi}\,v_\xi=v\,J_i^\pi\rangle$
changes its sign at $n_\xi=6$, where it vanishes.
The coefficient $\sum_\alpha c_{n_\xi \alpha}^2$ has the same sign
and the same order of magnitude between $n_\xi$ and $12-n_\xi$,
causing a large cancellation.
As a result, the $E2$ strength is significantly hindered for $^{152}$Yb.
Although the single-$j$ picture discussed by Lawson~\cite{ref:Lawson}
does not apply anymore, this mechanism,
$\langle N_{0h_{11/2}}\rangle\simeq 6$ or in other words
the cancellation of the matrix elements,
explains why the $E2$ hindrance occurs in $^{152}$Yb.
Thus the $10^+$ state of $^{152}$Yb yields a typical example
of the extended seniority isomerism.

The $E2$ strengths in the other even-$Z$ isotones
are also in remarkably good agreement with the data.
We clearly view improvement over the single-$j$ model in $^{154}$Hf.

As is viewed in Fig.~\ref{fig:Eng_e+},
the $4^+$ and $6^+$ states have not yet been detected
in $^{152}$Yb and $^{154}$Hf.
The ExSRF (\ref{eq:ExSRF}) approximately applies also
to the $8^+\rightarrow 6^+$ and $6^+\rightarrow 4^+$ $E2$ transitions.
These transitions are hindered by the same mechanism
as in the $10^+\rightarrow 8^+$ transition.
Hence it is not easy to populate the $4^+$ and $6^+$ states
in the experiments.

\begin{figure}[h]
\epsfysize=8cm
\centerline{\epsffile{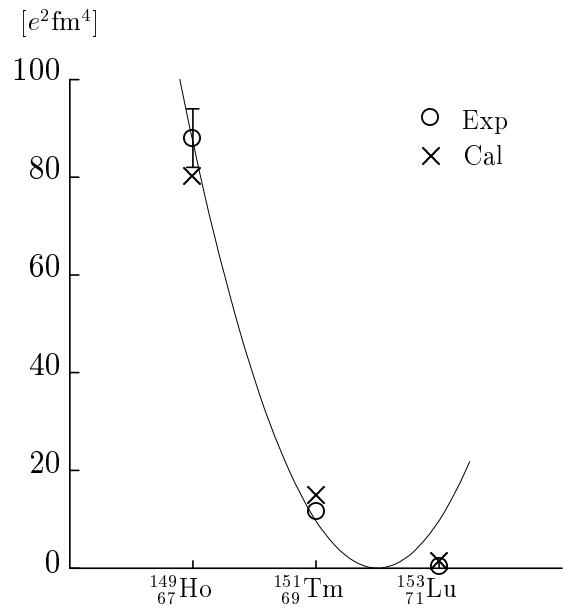}}
  \vspace{1cm}
  \caption{$B(E2;27/2^-\rightarrow 23/2^-)$ for odd-$Z$, $N=82$ nuclei.
  See Fig.~\protect\ref{fig:E2_e} for symbols.
  The experimental data are taken from
  Ref.~\protect\cite{ref:exp2} for $^{149}$Ho and $^{151}$Tm,
  and from Ref.~\protect\cite{ref:exp3} for $^{153}$Lu.
  }\label{fig:E2_o}
\end{figure}

The $E2$ strengths of the $27/2^-$ states
are shown in Fig.~\ref{fig:E2_o},
for the $67\leq Z\leq71, N=82$ isotones.
As in the $10^+$ isomers in the even-$Z$ isotones,
the present calculation reproduces the measured values
remarkably well.
The hindrance at $Z=71$ ({\it i.e.} $^{153}$Lu),
which was not described well in the single-$j$ model, is reproduced.
In the light of the ExSRF, this hindrance occurs
because of $\langle N_{0h_{11/2}}\rangle=6.2$ in $^{153}$Lu.
It should be emphasized that the $E2$ properties of the isomers
are naturally reproduced, by adjusting the energies
relevant to the excitation out of the $Z=64$ core.

\begin{figure}[h]
\epsfysize=8cm
\centerline{\epsffile{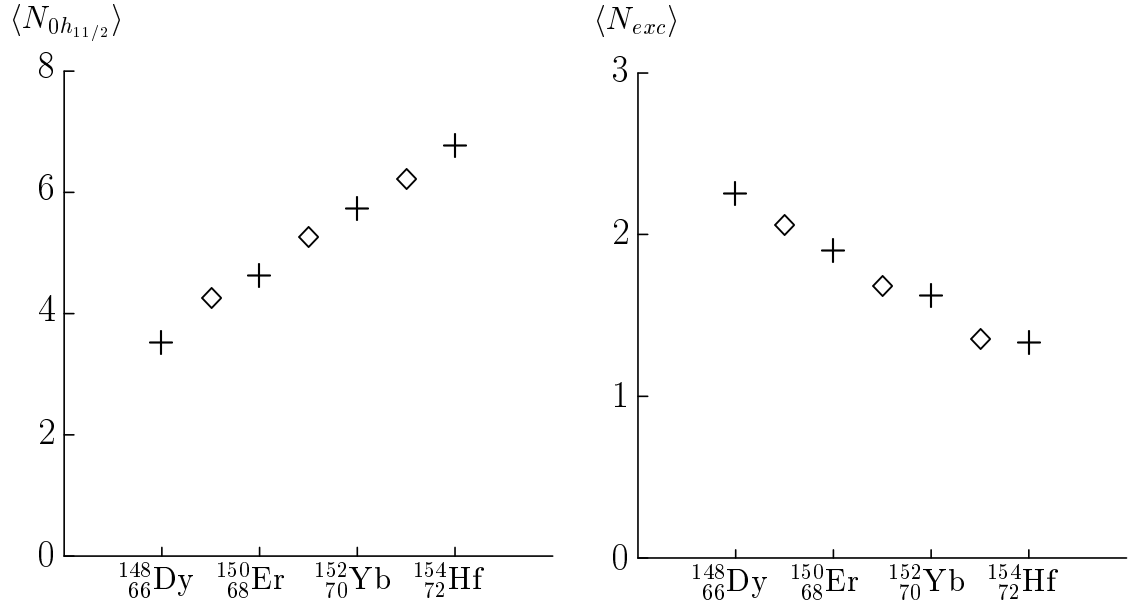}}
  \vspace{1cm}
  \caption{Left: occupation numbers $\langle N_{0h_{11/2}}\rangle$.
  Right: numbers of particles excited out of the $Z=64$ core,
  $\langle N_{\rm exc}\rangle \equiv
  14-(\langle N_{0g_{7/2}}\rangle + \langle N_{1d_{5/2}}\rangle)$.
  The plus symbols show the expectation values for the $10^+$ states,
  and the diamonds those for the $27/2^-$ states.
  }\label{fig:number}
\end{figure}

Figure~\ref{fig:number} depicts the occupation number
$\langle N_{0h_{11/2}}\rangle$ in the $10^+$ or $27/2^-$ isomers,
which corresponds to their $E2$ decay strengths via the ExSRF.
We view almost linear increase of $\langle N_{0h_{11/2}}\rangle$
in the isomers,
in coincidence with schematic illustration by Blomqvist
(Fig.~3-2 of Ref.~\cite{ref:Blom84}).
The number of the particles excited out of the $Z=64$ core
$\langle N_{\rm exc}\rangle \equiv
14-(\langle N_{0g_{7/2}}\rangle + \langle N_{1d_{5/2}}\rangle)$
in the isomers is plotted as well,
in the right panel of Fig.~\ref{fig:number}.
It is found that the number of the excited particles
diminishes only gradually, as $Z$ increases.

\section{Discussion --- Necessity of $Z=64$ core excitation}
\label{sec:discuss}

The ExSRF derived in Section~\ref{sec:ExSRF}
accounts for the seniority mechanism
to hinder the $E2$ decay of a certain class of isomers.
In Section~\ref{sec:multi-j}, we have demonstrated that
the strong $E2$ hindrance in $^{152}$Yb is reproduced
by taking the $Z=64$ core excitation into account.
However, the ExSRF itself does not exclude the possibility
of the single-$j$ solution to the hindrance for $^{152}$Yb.
In this section we argue that the $E2$ properties of the isomers
exclusively indicate the presence of the excitation across $Z=64$.

According to the ExSRF (\ref{eq:ExSRF}),
the vanishing $E2$ strength at $^{152}$Yb indicates
$\langle N_{0h_{11/2}}\rangle\simeq 6$.
In respect to the stiffness of the $Z=64$ core,
the following two possibilities result:
(i) $\pi 0h_{11/2}$ couples to the surrounding orbits
very weakly and the single-$j$ picture holds
to a good approximation, or
(ii) the pair excitation across $Z=64$ compensates
the pairing mixing of $0h_{11/2}$ with $(2s_{1/2}1d_{3/2})$,
the possibility first suggested by Blomqvist~\cite{ref:Blom84}.
We have shown in Section~\ref{sec:multi-j} that (ii) is plausible,
by reproducing the energy levels and the $E2$ strengths
simultaneously.
We here discuss whether (i) is possible or not.

For this purpose we consider the $E2$ strength of $^{152}$Yb
in the $3j$ model of $0h_{11/2}$, $2s_{1/2}$ and $1d_{3/2}$,
keeping the $Z=64$ closure.
The major point will be the amount of mixing of $0h_{11/2}$
with the surrounding orbits $2s_{1/2}$ and $1d_{3/2}$
due to the pairing interaction.
The possibility (i) requires that the mixing should be negligibly small.
The valence particle number $n$ is 6 for $^{152}$Yb
in the $3j$ model.
In order for the strong hindrance to be reproduced,
the wave function of $^{152}$Yb should have $n_\xi=6$
($\xi=0h_{11/2}$) as the main component,
with a small admixture of the $n_\xi=4$ component.
By using the effective charge of Ref.~\cite{ref:Lawson},
the measured $B(E2)$ leads to the admixture
of the $n_\xi=4$ component by no greater than 10\%.

For the sake of simplicity, let us first consider
mixing between two configurations.
In reality, this mixing could be either of the $0h_{11/2}$-$2s_{1/2}$
or the $0h_{11/2}$-$1d_{3/2}$ pairing mixing.
The degree of the mixing is connected to the ratio
of the off-diagonal matrix elements
of the pairing interaction
(denoted by $\langle V^{\rm off}_{\rm pair}\rangle$)
to the energy difference of the relevant orbits
(denoted by $\Delta E$).
The mixing probability is given by
$\langle V^{\rm off}_{\rm pair}\rangle^2
/[(2\Delta E)^2 + \langle V^{\rm off}_{\rm pair}\rangle^2]$.
The above 10\% mixing indicates $\langle V^{\rm off}_{\rm pair}\rangle
/2\Delta E = 0.45$.
If the mixing among the three orbits is considered,
this ratio should be regarded as the upper limit
for each of the $0h_{11/2}$-$2s_{1/2}$
and the $0h_{11/2}$-$1d_{3/2}$ mixing.
The level scheme of the odd-$Z$ isotones $^{147}$Tb,
$^{149}$Ho and $^{151}$Lu implies that
the three orbits keep nearly degenerate within the 0.2~MeV accuracy
in this region.
Hence we can put $\Delta E < 0.2$~MeV.
We thus find that, in order for the possibility (i) to be realized,
$\langle V^{\rm off}_{\rm pair}\rangle < 0.18$~MeV is necessary.

Generally speaking, interactions with the shorter range yield
the larger off-diagonal pairing matrix elements.
We estimate $\langle V^{\rm off}_{\rm pair}\rangle$ in the SDI
and in the Yukawa form with the range of the one-pion exchange,
as representatives of short-range and long-range interactions.
After fitting their strengths to the observed energy levels
in the $3j$ model,
the SDI and the Yukawa interaction give
$\langle V^{\rm off}_{\rm pair}\rangle\sim 1$ and 0.5~MeV,
respectively~\cite{ref:Mat99}.
As a consequence of the weak coupling between $0h_{11/2}$
and $(2s_{1/2}1d_{3/2})$,
the long-range interaction gives very low $0^+_2$ states,
enough low to be first excited states,
in $^{148}$Dy and $^{150}$Er.
Such $0^+_2$ levels have not been observed.
Nevertheless, even the long-range interaction
gives significantly larger $\langle V^{\rm off}_{\rm pair}\rangle$
than required in (i).
It is practically impossible to avoid a considerable mixing
of $(2s_{1/2}1d_{3/2})$ due to the pairing interaction.
Indeed, in both calculations with the SDI and the Yukawa interaction
in the $3j$ model space,
the three orbits well mix one another via the pairing interaction.
Thereby the $E2$ strengths of the high-spin isomers are almost
described by the formula (\ref{eq:GSRF}),
the QS formula for the degenerate single-particle orbits,
with $\Omega=9$.
This is obviously inconsistent with the measurement.

We thus conclude that the possibility (i) cannot be realistic.
As has been discussed in Section~\ref{sec:ExSRF},
in order to reproduce the $E2$ hindrance for $^{152}$Yb
the admixture of the $2s_{1/2}$ and $1d_{3/2}$ orbits
must be compensated by the $Z=64$ core excitation.
Therefore, with the nearly degeneracy
between $0h_{11/2}$ and $(2s_{1/2}1d_{3/2})$ taken into consideration,
the $E2$ properties of the $N=82$ isomers
are an exclusive evidence for the presence of the excitation
across $Z=64$,
not indicating stiff $Z=64$ core.
The seniority isomerism in this region
is a probe sensitive to the $Z=64$ core excitation
due to the pairing correlations.
It is difficult to handle the influence of the $Z=64$ core breaking
on the $E2$ properties of the isomers by renormalization.
The presence of substantial pair excitation
is a clear difference of the submagic number $Z=64$
from the ordinary magic numbers.

It is commented here that, in contrast to the SDI
adopted in Section~\ref{sec:multi-j},
Wenes {\it et al.} applied a finite-range interaction
with the Gaussian form to the nuclei in this region~\cite{ref:WHWV}.
As a result of the weak coupling between $0h_{11/2}$
and $(2s_{1/2}1d_{3/2})$,
their calculation predicted $0^+_2$ states at unusually low energies
in $^{148}$Dy and $^{150}$Er,
quite similar to the above $3j$ case with the Yukawa interaction.

Wildenthal proposed an effective Hamiltonian
for the $N=82$ isotones~\cite{ref:Wil91}.
Starting from the SDI+QQ interaction,
the interaction matrix elements are fitted to the $50<Z\leq 72$ nuclei.
%Similar truncation to Section~\ref{sec:multi-j}
%was assumed for the $Z\geq 64$ nuclei.
While Wildenthal's Hamiltonian (with the assumed truncation)
nicely describes the energy levels,
it does not reproduce the $Z$-dependence of the $E2$ properties
of the isomers; in particular, the strong hindrance around $^{152}$Yb.
This is because the pair excitation out of the $Z=64$ core
is too small, at least for the high-spin isomers.
%In other words, despite the good agreement
%with most observed energy levels,
%Wildenthal's model has a problem in the pairing properties,
%to which the $E2$ strengths of the isomers are quite sensitive.

The above possibilities (i) and (ii) can also be judged
by future experiments.
Though the ExSRF (\ref{eq:ExSRF}) connects the $E2$ strength
to the occupation number $\langle N_{0h_{11/2}}\rangle$,
the ambiguity in $e_{\rm eff}$ prohibits us
from extracting $\langle N_{0h_{11/2}}\rangle$
directly from $B(E2)$ of the isomers, in practice.
The two possibilities (i) and (ii) give
somewhat similar $E2$ strengths for the high-spin isomers
on account of the difference in $e_{\rm eff}$,
in the discussions so far.
However, this does not apply to the transition
from the lowest $2^+$ state to the ground $0^+$ state.
In Table~\ref{tab:E2_20}, the $B(E2;2^+\rightarrow 0^+)$ values
calculated in the multi-$j$ model are compared with
those in the single-$j$ model.
Without state-dependence of the effective charges,
we predict $2.5\sim 5$ times larger $E2$ strengths
in the multi-$j$ space ({\it i.e.} (ii))
than in the single-$j$ space ({\it i.e.} (i)).
In the multi-$j$ model of Section~\ref{sec:multi-j},
there seems to be a problem with respect to the quadrupole collectivity.
Hence we should not expect an excellent precision
on the multi-$j$ prediction;
indeed, by a slight variation of the interaction
$B(E2;2^+\rightarrow 0^+)$ can deviate by 30\%
without influencing the $E2$ strengths of the isomers.
Still the big difference between the single-$j$ and multi-$j$ models
would enable us to judge which of the two possibilities
(i) or (ii) is reliable.

\section{Summary and outlook}
\label{sec:summary}

We have investigated the $10^+$ and $27/2^-$ isomers
of the $Z>64$, $N=82$ nuclei.
The extended seniority reduction formula has been derived
for the $E2$ decay strengths of the isomers,
under reasonable assumptions.
This formula links the $E2$ strength to the occupation number
on the $\pi 0h_{11/2}$ orbit,
apart from the ambiguity in the effective charge.
The extended formula accounts for the mechanism of the $E2$ hindrance,
which we have called extended seniority isomerism.

By taking into account the excitations from $(0g_{7/2}1d_{5/2})$
to $(0h_{11/2}2s_{1/2}1d_{3/2})$,
the binding energies, the energy levels of both parities
and the $B(E2)$ values have simultaneously been reproduced
in a multi-$j$ shell model calculation with the MSDI.
The $E2$ hindrance in $^{153}$Lu as well as in $^{152}$Yb
has been described quite well.
Combined with the approximate degeneracy among the $0h_{11/2}$,
$2s_{1/2}$ and $1d_{3/2}$ orbits,
the strong $E2$ hindrance around $Z=70$ exclusively indicates
the presence of the pair excitation out of the $Z=64$ core.
Thus the $Z=64$ core is not very stiff.
It is not always justified to assume the $^{146}$Gd inert core,
even for the relatively low-lying states in the $N=82$ isotones.
In this respect, the number $Z=64$ should
be distinguished from the magic numbers like $N=82$,
though it could be fair to be called {\it submagic} number.

The extended seniority isomerism may exist
in other single-closed nuclei and their neighbors.
While we have restricted our discussion to the $Z>64$, $N=82$ nuclei,
focusing on the stiffness of the $Z=64$ core,
it is of interest to apply similar approaches
to nuclei in other mass region.
Work in this line is under progress.

~\\
This work was supported in part by the Ministry of Education,
Science, Sports and Culture of Japan (Grant-in-Aid for Encouragement
of Young Scientists, No. 11740137).

\section*{Appendix: Argument on the condition (\lowercase{c})
in Section~III}%\ref{sec:ExSRF}}
\setcounter{subsection}{0}

The condition (c) in Section~\ref{sec:ExSRF} is expected
to be a good approximation.
Indeed, it is exactly derived from (a) and (b)
if the interaction within the $\xi$-subspace
has the zero-range character.
We prove it in this Appendix.

\subsection{General argument}

With assuming the conditions (a) and (b),
let us consider matrix elements of the general shell model Hamiltonian
between the bases appearing
in Eqs.~(\ref{eq:state-i}) and (\ref{eq:state-f}).
We shall first prove the relation
\begin{eqnarray}
&&\left(\langle \xi^{n_\xi'}\,v_\xi=v\,J_f^\pi|\,
\langle r^{n-n_\xi'}\,\alpha'\,v_r=0\ 0^+|\right) H
\left(|\xi^{n_\xi}\,v_\xi=v\,J_f^\pi\rangle\,
|r^{n-n_\xi}\,\alpha\,v_r=0\ 0^+\rangle\right) \nonumber\\
&&- \left(\langle \xi^{n_\xi'}\,v_\xi=v\,J_i^\pi|\,
\langle r^{n-n_\xi'}\,\alpha'\,v_r=0\ 0^+|\right) H
\left(|\xi^{n_\xi}\,v_\xi=v\,J_i^\pi\rangle\,
|r^{n-n_\xi}\,\alpha\,v_r=0\ 0^+\rangle\right) \nonumber\\
&&= \delta_{n_\xi,n_\xi'}\delta_{\alpha,\alpha'}
\left(\langle \xi^{n_\xi}\,v\,J_f^\pi| V_\xi
|\xi^{n_\xi}\,v\,J_f^\pi\rangle -
\langle \xi^{n_\xi}\,v\,J_i^\pi| V_\xi
|\xi^{n_\xi}\,v\,J_i^\pi\rangle\right)\,,
\label{eq:matrix}
\end{eqnarray}
as far as the Hamiltonian consists of single-particle energies
and of two-body residual interaction.
Here $V_\xi$ stands for the two-body interaction within the $\xi$ subspace.
The left-hand side (LHS) of Eq.~(\ref{eq:matrix}) obviously
vanishes for the single-particle energy term of $H$.
It is sufficient to focus on matrix elements
of the two-body interaction.

The two-body interaction is expressed,
in the second-quantized form,
by the sum of the terms composed of
$a_{j_1}^\dagger a_{j_2}^\dagger a_{j_3} a_{j_4}$
operators (with coupling constants).
According to which of the $j$'s belong to $\xi$ or $r$,
all the possible terms contributing to the matrix elements
are classified into the following categories:
(i) $a_\xi^\dagger a_\xi^\dagger a_r a_r$ terms
and their hermitian conjugates,
(ii) $a_r^\dagger a_r^\dagger a_r a_r$ terms,
(iii) $a_\xi^\dagger a_r^\dagger a_r a_\xi$ terms,
and (iv) $a_\xi^\dagger a_\xi^\dagger a_\xi a_\xi$ terms.
It is noted that, since $v_r=0$,
the terms having an odd number of $(a_r^\dagger,a_r)$ operators vanish.
We decompose the operators into the $\xi$ part and the $r$ part,
and denote them by $\hat h_\xi$ and $\hat h_r$.
The matrix elements are also decomposed
into the $\xi$ and $r$ part,
\begin{eqnarray}
&&\left(\langle \xi^{n_\xi'}\,v_\xi=v\,J^\pi|\,
\langle r^{n-n_\xi'}\,\alpha'\,v_r=0\ 0^+|\right)
\hat h_\xi \hat h_r
\left(|\xi^{n_\xi}\,v_\xi=v\,J^\pi\rangle\,
|r^{n-n_\xi}\,\alpha\,v_r=0\ 0^+\rangle\right) \nonumber\\
&&= \langle \xi^{n_\xi'}\,v_\xi=v\,J^\pi|
\hat h_\xi |\xi^{n_\xi}\,v_\xi=v\,J^\pi\rangle\,
\langle r^{n-n_\xi'}\,\alpha'\,v_r=0\ 0^+|
\hat h_r |r^{n-n_\xi}\,\alpha\,v_r=0\ 0^+\rangle\,.
\end{eqnarray}
It is obvious that the $\langle r^{n-n_\xi'}\,\alpha'\,v_r=0\ 0^+|
\hat h_r |r^{n-n_\xi}\,\alpha\,v_r=0\ 0^+\rangle$ part
does not depend on $J^\pi$.
Therefore,
\begin{eqnarray}
&&\left(\langle \xi^{n_\xi'}\,v_\xi=v\,J_f^\pi|\,
\langle r^{n-n_\xi'}\,\alpha'\,v_r=0\ 0^+|\right) \hat h_\xi \hat h_r
\left(|\xi^{n_\xi}\,v_\xi=v\,J_f^\pi\rangle\,
|r^{n-n_\xi}\,\alpha\,v_r=0\ 0^+\rangle\right) \nonumber\\
&&- \left(\langle \xi^{n_\xi'}\,v_\xi=v\,J_i^\pi|\,
\langle r^{n-n_\xi'}\,\alpha'\,v_r=0\ 0^+|\right) \hat h_\xi \hat h_r
\left(|\xi^{n_\xi}\,v_\xi=v\,J_i^\pi\rangle\,
|r^{n-n_\xi}\,\alpha\,v_r=0\ 0^+\rangle\right) \nonumber\\
&&= \langle r^{n-n_\xi'}\,\alpha'\,v_r=0\ 0^+|
\hat h_r |r^{n-n_\xi}\,\alpha\,v_r=0\ 0^+\rangle \nonumber\\
&&\quad\quad\times \left(\langle \xi^{n_\xi'}\,v_\xi=v\,J_f^\pi|
\hat h_\xi |\xi^{n_\xi}\,v_\xi=v\,J_f^\pi\rangle -
\langle \xi^{n_\xi'}\,v_\xi=v\,J_i^\pi| \hat h_\xi
  |\xi^{n_\xi}\,v_\xi=v\,J_i^\pi\rangle\right)\,.
\label{eq:matrix-1}
\end{eqnarray}

For the matrix elements between the $v_r=0$ bases,
$\hat h_r$ cannot carry angular momentum,
and therefore $\hat h_\xi$ cannot either.
Then $\hat h_\xi$ and $\hat h_r$ for each category are defined as,
without loss of generality,
(i) $\hat h_\xi=[a_\xi^\dagger a_\xi^\dagger]^{(0)}$,
$\hat h_r=[\tilde a_r \tilde a_r]^{(0)}$,
(ii) $\hat h_\xi=1$, $\hat h_r=[a_r^\dagger a_r^\dagger
\tilde a_r \tilde a_r]^{(0)}$,
(iii) $\hat h_\xi=[a_\xi^\dagger \tilde a_\xi]^{(0)}$,
$\hat h_r=[a_r^\dagger \tilde a_r]^{(0)}$,
and (iv) $\hat h_\xi=[a_\xi^\dagger a_\xi^\dagger
\tilde a_\xi \tilde a_\xi]^{(0)}$, $\hat h_r=1$.
$V_\xi$ in Eq.~(\ref{eq:matrix}) represents the collection
of the $\hat h_\xi$'s belonging to (iv).
We discuss the matrix elements of $\hat h_\xi$
in the right-hand side (RHS) of Eq.~(\ref{eq:matrix-1}),
respective to the above four categories.

The category (i) leads to
$n_\xi'=n_\xi\pm 2$ off-diagonal elements.
The $\hat h_\xi=[a_\xi^\dagger a_\xi^\dagger]^{(0)}$ operator
in this case is proportional to a generator
of the quasi-spin in the orbit $\xi$.
Thus $\langle \xi^{n_\xi+2}\,v_\xi=v\,J^\pi|
\hat h_\xi |\xi^{n_\xi}\,v_\xi=v\,J^\pi\rangle$
depends only on $n_\xi$ and $v_\xi$, not on $J^\pi$.
Namely,
\begin{equation}
\langle \xi^{n_\xi'}\,v_\xi=v\,J_f^\pi| [a_\xi^\dagger a_\xi^\dagger]^{(0)}
  |\xi^{n_\xi}\,v_\xi=v\,J_f^\pi\rangle =
\langle \xi^{n_\xi'}\,v_\xi=v\,J_i^\pi| [a_\xi^\dagger a_\xi^\dagger]^{(0)}
  |\xi^{n_\xi}\,v_\xi=v\,J_i^\pi\rangle\,, \label{eq:ximat-1}
\end{equation}
and the RHS of Eq.~(\ref{eq:matrix-1}) vanishes.

Since $\hat h_\xi=1$ in the category (ii),
the relevant matrix elements of the $\xi$ part are
\begin{equation}
\langle \xi^{n_\xi'}\,v_\xi=v\,J_f^\pi|\xi^{n_\xi}\,v_\xi=v\,J_f^\pi\rangle =
\langle \xi^{n_\xi'}\,v_\xi=v\,J_i^\pi|\xi^{n_\xi}\,v_\xi=v\,J_i^\pi\rangle
= \delta_{n_\xi,n_\xi'}\,. \label{eq:ximat-2}
\end{equation}
For the category (iii), $\hat h_\xi=[a_\xi^\dagger \tilde a_\xi]^{(0)}
\propto N_\xi$, leading to
\begin{equation}
\langle \xi^{n_\xi'}\,v_\xi=v\,J_f^\pi| [a_\xi^\dagger \tilde a_\xi]^{(0)}
  |\xi^{n_\xi}\,v_\xi=v\,J_f^\pi\rangle =
\langle \xi^{n_\xi'}\,v_\xi=v\,J_i^\pi| [a_\xi^\dagger \tilde a_\xi]^{(0)}
  |\xi^{n_\xi}\,v_\xi=v\,J_i^\pi\rangle
\propto \delta_{n_\xi,n_\xi'}n_\xi\,. \label{eq:ximat-3}
\end{equation}
Both terms do not contribute to the RHS of Eq.~(\ref{eq:matrix-1}).

 From Eqs.~(\ref{eq:ximat-1}), (\ref{eq:ximat-2}) and (\ref{eq:ximat-3}),
only the terms of the category (iv) may contribute
to the RHS of Eq.~(\ref{eq:matrix-1}).
Equation~(\ref{eq:matrix}) follows, because $\hat h_r=1$ for the category (iv),
with replacing the sum of $\hat h_\xi$'s by $V_\xi$.
The argument is now reduced to the single-$j$ matrix elements
within the $\xi$ orbit.
Remark again that we have not imposed any restriction on the Hamiltonian
in the discussion so far,
besides that it consists of the single-particle energies
and two-body interaction.

\subsection{Property of $V_\xi$}

We next consider the property of $V_\xi$.
If $\langle \xi^{n_\xi}\,v\,J^\pi| V_\xi |\xi^{n_\xi}\,v\,J^\pi\rangle$
is independent of $n_\xi$ or $J$,
we have
\begin{eqnarray}
&&\langle \xi^{n_\xi}\,v\,J_f^\pi| V_\xi |\xi^{n_\xi}\,v\,J_f^\pi\rangle -
\langle \xi^{n_\xi}\,v\,J_i^\pi| V_\xi |\xi^{n_\xi}\,v\,J_i^\pi\rangle
\nonumber\\
&&= \langle \xi^v\,v\,J_f^\pi| V_\xi |\xi^v\,v\,J_f^\pi\rangle -
\langle \xi^v\,v\,J_i^\pi| V_\xi |\xi^v\,v\,J_i^\pi\rangle\,.
\label{eq:ximat-4}
\end{eqnarray}
Substituting it into the RHS of Eq.~(\ref{eq:matrix}),
we obtain
\begin{eqnarray}
&&\left(\langle \xi^{n_\xi'}\,v_\xi=v\,J_f^\pi|\,
\langle r^{n-n_\xi'}\,\alpha'\,v_r=0\ 0^+|\right) H
\left(|\xi^{n_\xi}\,v_\xi=v\,J_f^\pi\rangle\,
|r^{n-n_\xi}\,\alpha\,v_r=0\ 0^+\rangle\right) \nonumber\\
&&- \left(\langle \xi^{n_\xi'}\,v_\xi=v\,J_i^\pi|\,
\langle r^{n-n_\xi'}\,\alpha'\,v_r=0\ 0^+|\right) H
\left(|\xi^{n_\xi}\,v_\xi=v\,J_i^\pi\rangle\,
|r^{n-n_\xi}\,\alpha\,v_r=0\ 0^+\rangle\right) \nonumber\\
&&= \delta_{n_\xi,n_\xi'}\delta_{\alpha,\alpha'}
\left(\langle \xi^v\,v\,J_f^\pi| V_\xi |\xi^v\,v\,J_f^\pi\rangle -
\langle \xi^v\,v\,J_i^\pi| V_\xi |\xi^v\,v\,J_i^\pi\rangle\right)\,.
\label{eq:matrix-2}
\end{eqnarray}
The Hamiltonian matrix can be separated according to $J$,
because of the angular momentum conservation.
Moreover, since the seniority $v$ has been assumed to be
a good quantum number,
it is sufficient to consider submatrices of $H$
for a fixed $v$.
The space to be diagonalized is spanned
by $|\xi^{n_\xi}\,v_\xi=v\,J^\pi\rangle \,
|r^{n-n_\xi}\,\alpha\,v_r=0\ 0^+\rangle$
with various $n_\xi$ and $\alpha$
(see Eqs.~(\ref{eq:state-i}) and (\ref{eq:state-f})).
Equation~(\ref{eq:matrix-2}) implies that
the submatrices of $H$ are identical
between $|(\xi r)^n\,v\,J_i^\pi\rangle$ and $|(\xi r)^n\,v\,J_f^\pi\rangle$,
except for a constant shift of the diagonal elements.
Diagonalized by the same unitary matrix,
the lowest eigenstates $|J_i^\pi\rangle$ and $|J_f^\pi\rangle$
have equal coefficient $c_{n_\xi \alpha}$ to each other;
the condition (c) is exactly satisfied.
It is now clear that Eq.~(\ref{eq:ximat-4}) is crucial
to the condition (c).

In the quasi-spin (QS) regime within the single orbit $\xi$,
$[a_\xi^\dagger a_\xi^\dagger \tilde a_\xi \tilde a_\xi]^{(0)}$
can be QS-scalar, vector or tensor, in general.
If $V_\xi$ is purely QS-scalar,
the matrix element $\langle \xi^{n_\xi}\,v_\xi=v\,J_i^\pi| V_\xi
|\xi^{n_\xi}\,v_\xi=v\,J_i^\pi\rangle$ is independent of $n_\xi$,
and Eq.~(\ref{eq:ximat-4}) is fulfilled.
Equation~(\ref{eq:ximat-4}) is also satisfied
if QS-vector and QS-tensor parts of $V_\xi$ are $J$-independent.
This is indeed attained
when the QS-vector and tensor parts can be expressed
by the QS generators ($[a_\xi^\dagger a_\xi^\dagger]^{(0)}$,
$[\tilde a_\xi \tilde a_\xi]^{(0)}$ and
$[a_\xi^\dagger \tilde a_\xi]^{(0)}$,
besides appropriate constant factors).
An immediate example is the monopole pairing
($[a_\xi^\dagger a_\xi^\dagger]^{(0)}
[\tilde a_\xi \tilde a_\xi]^{(0)}$).
A sufficient condition to Eq.~(\ref{eq:ximat-4}) is
that $V_\xi$ consists only of QS-scalars and of the QS generators.

The general form of $V_\xi$ can be represented by
\begin{equation}
V_\xi = - \sum_{\lambda={\rm even}}
  {{g_\lambda}\over 2}\, [a_\xi^\dagger a_\xi^\dagger]^{(\lambda)}
\cdot[\tilde a_\xi \tilde a_\xi]^{(\lambda)}\,. \label{eq:Vxi}
\end{equation}
The corresponding `particle-hole' interaction
is defined by~\cite{ref:AI66}
\begin{equation}
\bar V_\xi = \sum_\lambda
  f_\lambda\, [a_\xi^\dagger a_\xi^\dagger]^{(\lambda)}
\cdot[\tilde a_\xi \tilde a_\xi]^{(\lambda)}\,, \label{eq:ph-int}
\end{equation}
with
\begin{equation}
f_\lambda = \sum_{\lambda'={\rm even}} (2\lambda'+1)\,
W(j_\xi j_\xi j_\xi j_\xi;\lambda' \lambda)\,g_{\lambda'}\,.
\end{equation}
In Eq.~(\ref{eq:ph-int}), only $\lambda={\rm even}$ terms remain
owing to the antisymmetrization.
According to the QS argument~\cite{ref:AI66}, we have
\begin{eqnarray}
\langle \xi^{n_\xi}\,v\,J^\pi| V_\xi |\xi^{n_\xi}\,v\,J^\pi\rangle
&=& \left\{ {{(\Omega_\xi-2v)(2\Omega_\xi-n_\xi-v)}\over
{4(\Omega_\xi-v)(\Omega_\xi-v-1)}} (g_0+2f_0) - f_0 \right\} (n_\xi-v)
\nonumber\\
&&+ {{(\Omega_\xi-v)(\Omega_\xi-v-2) + (n_\xi-\Omega_\xi)^2}\over
{2(\Omega_\xi-v)(\Omega_\xi-v-1)}}
\langle \xi^v\,v\,J^\pi| V_\xi |\xi^v\,v\,J^\pi\rangle \nonumber\\
&&+ {{(\Omega_\xi-v)^2 - (n_\xi-\Omega_\xi)^2}\over
{2(\Omega_\xi-v)(\Omega_\xi-v-1)}}
\langle \xi^v\,v\,J^\pi| \bar V_\xi |\xi^v\,v\,J^\pi\rangle \,.
\label{eq:matrix-3}
\end{eqnarray}
By subtracting out the $n_\xi$- and $J$-independent terms,
Eq.~(\ref{eq:matrix-3}) derives
\begin{eqnarray}
&&\left(\langle \xi^{n_\xi}\,v\,J_f^\pi| V_\xi
|\xi^{n_\xi}\,v\,J_f^\pi\rangle -
\langle \xi^{n_\xi}\,v\,J_i^\pi| V_\xi |\xi^{n_\xi}\,v\,J_i^\pi\rangle\right)
  \nonumber\\
&&\quad\quad - \left(\langle \xi^v\,v\,J_f^\pi| V_\xi
|\xi^v\,v\,J_f^\pi\rangle -
\langle \xi^v\,v\,J_i^\pi| V_\xi |\xi^v\,v\,J_i^\pi\rangle\right)
\nonumber\\
&&= {{(n_\xi-\Omega_\xi)^2}\over{2(\Omega_\xi-v)(\Omega_\xi-v-1)}}
\left\{\langle \xi^v\,v\,J_f^\pi| (V_\xi-\bar V_\xi)
|\xi^v\,v\,J_f^\pi\rangle -
\langle \xi^v\,v\,J_i^\pi| (V_\xi-\bar V_\xi)
|\xi^v\,v\,J_i^\pi\rangle\right\}\,.
\label{eq:ximat-5}
\end{eqnarray}
The $n_\xi$-dependence is eliminated if we have
\begin{equation}
\langle \xi^v\,v\,J_f^\pi| (V_\xi-\bar V_\xi) |\xi^v\,v\,J_f^\pi\rangle =
\langle \xi^v\,v\,J_i^\pi| (V_\xi-\bar V_\xi) |\xi^v\,v\,J_i^\pi\rangle\,.
\label{eq:ximat-6}
\end{equation}

Equations~(\ref{eq:Vxi}) and (\ref{eq:ph-int}) lead to
\begin{equation}
V_\xi - \bar V_\xi = - \sum_{\lambda={\rm even}}
{{g_\lambda + 2f_\lambda}\over 2}
\, [a_\xi^\dagger a_\xi^\dagger]^{(\lambda)}
\cdot[\tilde a_\xi \tilde a_\xi]^{(\lambda)}\,. \label{eq:Vred}
\end{equation}
As far as the seniority is not large,
only a limited number of $\lambda$'s in Eq.~(\ref{eq:Vred})
contribute to the RHS of Eq.~(\ref{eq:ximat-5}).
For instance, only the $\lambda=J_i$ and $J_f$ terms
are relevant to the $v=2$ case,
and Eq.~(\ref{eq:ximat-6}) then derives
$g_{J_i}+2f_{J_i} = g_{J_f}+2f_{J_f}$.
When $V_\xi$ is QS-scalar, $V_\xi=\bar V_\xi$ and
$g_\lambda+2f_\lambda=0$ for any even $\lambda$.

\subsection{Zero-range interaction}

We here verify that Eq.~(\ref{eq:ximat-4}) is exactly fulfilled
if $V_\xi$ is a zero-range interaction.

The one-body operator $[a_\xi^\dagger \tilde a_\xi]^{(\lambda)}$
is QS-scalar for an odd $\lambda$~\cite{ref:AI66}.
An easy way to construct a QS-scalar interaction
is to take
\begin{equation}
\sum_{\lambda={\rm odd}} q_\lambda\,
[a_\xi^\dagger \tilde a_\xi]^{(\lambda)}\cdot
[a_\xi^\dagger \tilde a_\xi]^{(\lambda)}\,,
\end{equation}
where $q_\lambda$ is an arbitrary constant.

Suppose that $V_\xi$ is a zero-range interaction,
which we here define as
\begin{equation}
V_\xi^S = u(r_1,r_2) \delta(\hat{\mathbf r}_1-\hat{\mathbf r}_2)\,,
\label{eq:zero}
\end{equation}
with the exchange symmetry $u(r_1,r_2)=u(r_2,r_1)$.
The SDI adopted in the text is of this type
($u(r_1,r_2)\propto\delta(r_1-r_2)\delta(r_1-R)/R^2$).
Since $V_\xi$ is under discussion,
the radial part of the interaction is unimportant,
giving only an overall factor to the matrix elements.
Expanding the angular part of $V_\xi^S$
by the Legendre polynomials, we obtain
\begin{eqnarray}
V_\xi^S &=& u(r_1,r_2) \sum_\lambda {{2\lambda+1}\over 2}
P_\lambda(\cos\theta_{12})\nonumber\\
&=& 2\pi\,u(r_1,r_2) \sum_\lambda
Y^{(\lambda)}(\hat{\mathbf r}_1)\cdot Y^{(\lambda)}(\hat{\mathbf r}_2)\,.
\end{eqnarray}
On the other hand, the zero-range interaction
given in Eq.~(\ref{eq:zero}) acts
on the spatially symmetric two-body states.
Therefore, if the two-body states are antisymmetrized,
the zero-range interaction automatically picks up
the spin-singlet two-body states for identical fermion systems.
This leads to~\cite{ref:Tal93}
\begin{equation}
V_\xi^S = -{4\over 3} ({\mathbf s}_1\cdot{\mathbf s}_2) V_\xi^S
= -{{8\pi}\over 3}\,u(r_1,r_2)\,({\mathbf s}_1\cdot{\mathbf s}_2)
\sum_\lambda
Y^{(\lambda)}(\hat{\mathbf r}_1)\cdot Y^{(\lambda)}(\hat{\mathbf r}_2)\,.
\end{equation}
With the angular momentum recoupling, we rewrite it as
\begin{equation}
V_\xi^S = -{{8\pi}\over 3}\,u(r_1,r_2) \sum_{\lambda,\kappa}
[Y^{(\lambda)}(\hat{\mathbf r}_1) {\mathbf s}_1]^{(\kappa)}
\cdot [Y^{(\lambda)}(\hat{\mathbf r}_2) {\mathbf s}_2]^{(\kappa)}\,.
\label{eq:Ys}
\end{equation}

We now switch to the second-quantized representation.
The equivalent one-body operator to
$[Y^{(\lambda)}(\hat{\mathbf r}) {\mathbf s}]^{(\kappa)}$
in the $\xi$ subspace is
\begin{equation}
{1\over{2\kappa+1}} \langle \xi||
[Y^{(\lambda)}(\hat{\mathbf r}) {\mathbf s}]^{(\kappa)} ||\xi\rangle
[a_\xi^\dagger \tilde a_\xi]^{(\kappa)}\,.
\label{eq:1body}
\end{equation}
The single-particle matrix element in Eq.~(\ref{eq:1body}) is
evaluated by
\begin{equation}
\langle \xi|| [Y^{(\lambda)}(\hat{\mathbf r}) {\mathbf s}]^{(\kappa)}
  ||\xi\rangle
= \sqrt{2\kappa+1}\,(2j_\xi+1) \left\{\begin{array}{ccc}
l_\xi& 1/2& j_\xi\\ \lambda& 1& \kappa\\
l_\xi& 1/2& j_\xi \end{array}\right\}
\langle l_\xi||Y^{(\lambda)}(\hat{\mathbf r})||l_\xi\rangle\,
\langle 1/2||{\mathbf s}||1/2\rangle\,.
\label{eq:1b-mat}
\end{equation}
In order for
$\langle l_\xi||Y^{(\lambda)}(\hat{\mathbf r})||l_\xi\rangle$
in the RHS not to vanish,
$\lambda$ must be even (parity selection rule).
On the other hand, owing to the symmetry of the $9j$-symbol
in Eq.~(\ref{eq:1b-mat}),
the above matrix element vanishes if $\lambda+1+\kappa$ is odd.
This is a consequence of the time reversality.
Therefore, the single-particle matrix element of Eq.~(\ref{eq:1b-mat})
vanishes for even $\kappa$.
Back to Eq.~(\ref{eq:1body}),
we find that $[Y^{(\lambda)}(\hat{\mathbf r}) {\mathbf s}]^{(\kappa)}$
is QS-scalar because $\kappa$ is always odd.
Hence $V_\xi^S$ is also QS-scalar via Eq.~(\ref{eq:Ys}).
Thus, the zero-range $V_\xi$ has $g_\lambda + 2f_\lambda =0$
in Eq.~(\ref{eq:Vred}) for any possible $\lambda$,
and therefore satisfies Eq.~(\ref{eq:ximat-4}).

In reality, $V_\xi$ will not fully be zero-range.
However, as far as the short-range interaction dominates,
the matrix elements of $V_\xi$ are not very different
from the zero-range interaction,
having $g_\lambda + 2f_\lambda \simeq 0$.
The condition (c) is therefore expected to be a good approximation.

\clearpage
%%%%%%%%%%%%%%%%%%%%%%%%%%%%%%%%%%%%%%%%%%%%%%%%%%%%%%%%%%%%%%%%%%%%%%%%%%%%%
%%%%%%%%%%%%%%%%%%%%%%%%%%%%      TABLE     %%%%%%%%%%%%%%%%%%%%%%%%%%%%%%%%%
%%%%%%%%%%%%%%%%%%%%%%%%%%%%%%%%%%%%%%%%%%%%%%%%%%%%%%%%%%%%%%%%%%%%%%%%%%%%%
\begin{table}
\caption{Adopted values of the single particle energies.}\label{tab:spe}
\begin{tabular}{lrrrrr}
  $j$ & $0g_{7/2}$ & $1d_{5/2}$ & $0h_{11/2}$ & $2s_{1/2}$ &
  $1d_{3/2}$   \\ \hline
  $\epsilon_j$~(MeV) & $-8.33$ & $-7.73$ &  $-6.88$ & $-6.73$ & $-6.43$ \\
\end{tabular}
\end{table}

\begin{table}[h]
  \vspace{0cm}
  \caption{Contribution of each $n_\xi$ component
  ($\xi=0h_{11/2}$) to the $E2$ matrix element
  of the $10^+\rightarrow 8^+$ transition in $^{152}$Yb
  (see text and Eq.~(\protect\ref{eq:ExSRF0})).
  The third row shows the matrix element
  $\langle \xi^{n_\xi};8^+||T(E2)||\xi^{n_\xi};10^+\rangle/e_{\rm eff}$,
  which is evaluated by using the harmonic oscillator
  single-particle wave functions with $\nu=0.18~{\rm fm}^{-2}$.
  }\label{tab:wf_Yb}
  \begin{tabular}{lrrrrrr}
  $n_\xi$ & $2$ & $4$ & $6$ & $8$ & $10$ & Sum \\ \hline
  $\sum_\alpha c_{n_\xi \alpha}^2$
  & 0.028 & 0.282 & 0.498 & 0.181 & 0.010 & 1.00 \\
  Matrix element~(${\rm fm}^2$)
   & $21.9$ & $ 11.0$  & $0$  & $-11.0$ & $-21.9$ & --- \\
  Product~(${\rm fm}^2$) & 0.614 & 3.09 & 0 & $-1.98$ & $-0.219$ & 1.50 \\
\end{tabular}
\end{table}

\begin{table}
\caption{Calculated $B(E2;2^+\rightarrow 0^+)$ values ($e^2{\rm fm}^4$).
The $0h_{11/2}$ single-$j$ results using the parameters
of Ref.~\protect\cite{ref:Lawson} and those of the present work (PW)
are compared.}\label{tab:E2_20}
\begin{tabular}{cccccc}
    & $^{148}_{\ 66}$Dy   &  $^{150}_{\ 68}$Er  & $^{152}_{\ 70}$Yb
    &  $^{154}_{\ 72}$Hf  \\ \hline
  Ref.~\cite{ref:Lawson}  & 187 &  298 &  335 &   298 \\
  PW & 967 & 960 & 886 &  787 \\
\end{tabular}
\end{table}

\clearpage

%%%%%%%%%%%%%%%%%%%%%%%%%%%%%%%%%%%%%%%%%%%%%%%%%%%%%%%%%%%%%%%%%%%%%%%%%%%%%
%%%%%%%%%%%%%%%%%%%%%%%%%%%%     FIGURE     %%%%%%%%%%%%%%%%%%%%%%%%%%%%%%%%%
%%%%%%%%%%%%%%%%%%%%%%%%%%%%%%%%%%%%%%%%%%%%%%%%%%%%%%%%%%%%%%%%%%%%%%%%%%%%%

\clearpage

\clearpage

\clearpage

\clearpage

\clearpage

\clearpage

\begin{references}
\bibitem{uv-factor} Do not confuse this $v_\xi$ with the seniority.
\end{references}
\end{document}